\documentclass[twocolumn,prd,floatfix,preprintnumbers,a4paper,nofootinbib,superscriptaddress,groupedaddress]{revtex4-1}
\usepackage{natbib}
\usepackage[english]{babel}
\usepackage[autostyle, english = american]{csquotes}
\MakeOuterQuote{"}
\usepackage[utf8]{inputenc}
\usepackage{amsmath,amssymb}
\usepackage{amsfonts}
\usepackage{bm}
\usepackage{graphics}
\usepackage{siunitx}
\usepackage{float}
\usepackage{amsmath}
\usepackage{amssymb}
\usepackage[normalem]{ulem} 
\usepackage[mathscr]{euscript}
\usepackage{acronym}
\usepackage{makecell}
\usepackage{float}
\usepackage[caption=false]{subfig}
\usepackage{lipsum}
\usepackage{mathrsfs}
\usepackage{mathtools}
\usepackage{multirow}
\usepackage{graphicx}
\usepackage{mathtools}
\usepackage{multirow}
\usepackage{graphicx}
\usepackage{booktabs}
\usepackage[usenames,dvipsnames,table,xcdraw]{xcolor}
\usepackage[colorlinks, urlcolor=blue,citecolor=blue,pdfborder={0 0 0}]{hyperref} 
\usepackage[nameinlink,capitalise]{cleveref}
\usepackage{xspace}
\usepackage{xfp}
\usepackage{array}

\newcommand{\reply}[1]{{#1}}

\begin{document}

\title{Impact of Neutrino Flavor Conversions on Neutron Star Merger Dynamics, Ejecta, Nucleosynthesis, and Multi-Messenger Signals}

\author{Yi Qiu$^{1,2}$, David Radice$^{1,2,3}$, Sherwood Richers$^4$, Federico Maria Guercilena$^{5,6}$, Albino Perego$^{5,6}$ and Maitraya Bhattacharyya$^{1,2}$}

\affiliation{$^1$Institute for Gravitation and the Cosmos, The Pennsylvania State University, University Park PA 16802, USA}
\affiliation{$^2$Department of Physics, The Pennsylvania State University, University Park PA 16802, USA}
\affiliation{$^3$Department of Astronomy \& Astrophysics, The Pennsylvania State University, University Park PA 16802, USA}
\affiliation{$^4$Department of Physics \& Astronomy, University of Tennessee Knoxville, Knoxville TN 37996, USA}
\affiliation{$^5$Dipartimento di Fisica, Università di Trento, via Sommarive 14, 38123 Trento, Italy}
\affiliation{$^6$INFN-TIFPA, Trento Institute for Fundamental Physics and Applications, via Sommarive 14, 38123 Trento, Italy}

\date{\today}

\begin{abstract}
We present numerical relativity simulations of binary neutron star mergers incorporating neutrino flavor transformations triggered by fast flavor instability, quantum many-body effects, or potential beyond standard model physics. In both long-lived and short-lived remnant scenarios, neutrino flavor conversions modify species-dependent neutrino luminosities and mean energies, and drive the matter towards more neutron rich conditions. They produce up to $300\%$ more neutron rich ejecta and significantly boost the r-process yields, especially in low-density, near-equatorial outflows. We identify regions unstable to fast flavor instabilities and find that these instabilities persist despite flavor conversions. We further test the sensitivity to the equilibration timescale of the flavor conversions, finding that slower flavor conversions can interact with thermodynamic equilibration, and increase the neutron richness of the ejecta. Flavor conversions may also contribute to stronger gravitational wave and neutrino emissions, pointing to a correlation between neutrino transport and merger dynamics. These results highlight the potential impact of flavor conversions while motivating future work to improve on theoretical understanding of flavor instabilities in global simulations.
\end{abstract}

\maketitle

\section{Introduction} \label{sec:intro}

Next-generation facilities for gravitational waves (GWs) and electromagnetic observations will enable multi-messenger studies of compact object mergers with unprecedented precision and across broader frequency bands~\cite{LIGOScientific:2016aoc, LIGOScientific:2017vwq, LIGOScientific:2017zic, KAGRA:2013rdx, Eichler:1989ve, Radice:2020ddv, Perego:2017wtu, Ruffert:1998qg, Just:2015dba, Cusinato:2021zin, Martin:2017dhc, Diamond:2023cto, Vigna-Gomez:2023euq,Gupta:2023lga,Niu:2025nha}. The upcoming observations can shape our understanding of some of the most fundamental questions in modern physics and astrophysics, like the nature of dense nuclear matter in stars~\cite{Lattimer:2004pg,Du:2021rhq,Burgio:2021vgk,Huth:2021bsp,Huang:2022mqp,Chatziioannou:2024tjq,Ecker:2024uqv,Most:2025kqf,Hammond:2025kki} and rapid neutron-capture (r-process) nucleosynthesis~\cite{Lattimer:1977igd,Meyer:1989ApJ,Qian:1996xt, Hoffman:1996aj, Freiburghaus:1999ApJ, Metzger:2010sy, Goriely:2011ApJ,Just:2014fka,Wu:2016pnw, Drout:2017ijr, Balantekin:2023ayx,Chen:2024gwj}. One of the most fascinating events to expect are binary neutron star (BNS) mergers, where the majority of r-process might take place. Currently, the exact compositions and yields of the mass ejection of merger remnants remain uncertain~\cite{Freiburghaus:1999ApJ,Korobkin:2012uy,Wanajo:2014wha, Thielemann:2017acv, Radice:2018pdn,Shibata:2019wef,Ekanger:2023mde,Ricigliano:2024lwf,Chen:2024acv,Ma:2025ouj,Jacobi:2025eak} both observationally and theoretically due to the complexity of the interplay between the merger dynamics, nuclear physics, atomic physics, and neutrino interactions~\cite{Espino:2023dei, Espino:2023mda, Espino:2022mtb,Pajkos:2024iry,Ng:2024zve,Foucart:2024npn,Gieg:2024jxs}. Theoretical modeling is essential to quantify these uncertainties and to interpret observations, requiring numerical solutions of Einstein’s equations of general relativity with matter.

Numerical simulations of BNS mergers generally attempt to evolve the full system of general relativistic hydrodynamics (GRHD) equations, augmented with sophisticated microphysics and neutrino radiation transport~\cite{Ruffert:1998qg,Rosswog:1998hy, Sekiguchi:2011zd,Sekiguchi:2015dma,Palenzuela:2015dqa,Foucart:2016rxm,Radice:2016dwd,Foucart:2022bth, Volpe:2023met, Kawaguchi:2024naa}. In particular, neutrino transport is the main energy transfer channel for BNS mergers, affecting both their dynamics and associated nucleosynthesis. Classically, the evolution of the neutrino radiation field in phase space is described by the Boltzmann equation. However, the prohibitive computational demand of direct Boltzmann solvers renders them infeasible for global-scale simulations~\cite{Sumiyoshi:2020bdh, Bhattacharyya:2022bzf}. As compromises, numerical codes typically employ either approximate moment-based transport schemes~\cite{Thorne:1981nvt, Shibata:2011kx, Cardall:2013kwa, Foucart:2016rxm, Sekiguchi:2015dma, Sekiguchi:2016bjd, Radice:2021jtw, Foucart:2024npn, Cheong:2024buu} or Monte Carlo methods~\cite{Foucart:2017mbt, Foucart:2025nub}. Hybrid strategies—such as the guided moments method~\cite{Izquierdo:2023fub}—have been proposed but not yet applied to BNS mergers.

Recently, one emerging concern with BNS simulations with neutrino transports is their neglecting of neutrino flavor conversions~\cite{Sigl:1993ctk, Sawyer:2005jk, Duan:2005cp, Izaguirre:2016gsx,Zhu:2016mwa, Chakraborty:2016lct, Capozzi:2017gqd, Wu:2017drk, Richers:2021xtf, Morinaga:2021vmc, Wu:2021uvt,Nagakura:2021hyb, Richers:2022zug, Capozzi:2022slf, Volpe:2023met, Abbar:2023zkm, Zaizen:2023wht, Johns:2023jjt, Johns:2024dbe, Johns:2025mlm}, which are prevalent in neutrino-dense environments. The electron fraction ($Y_e$) of the merger ejecta—a key parameter determining the final nucleosynthetic yields~\cite{Hoffman:1996aj,Lippuner:2015gwa}—is strongly influenced by the local equilibrium densities of electron-type neutrinos, which may vary when flavor oscillations persist. 

Early studies identified the collective neutrino flavor conversions in dense neutrino gases, see also~\cite{Duan:2010bg,Capozzi:2022slf,Volpe:2023met,Johns:2025mlm} for reviews. In particular, the fast flavor instabilities (FFI), a specific class of collective oscillations, is usually studied using idealized radiation-only models~\cite{Sawyer:2005jk,Richers:2021xtf, Xiong:2023vcm, Fiorillo:2024qbl, Xiong:2024pue, Kost:2024esc, Kneller:2024buy, Lacroix:2024pbb,Xiong:2024pue, Nagakura:2025brr,Fiorillo:2025zio,Johns:2025yxa,Laraib:2025uza,Fiorillo:2025kko,Fiorillo:2025npi}. Some more recent works start to extend these analyses to static snapshots from classical simulations~\cite{Wu:2017drk, Richers:2022dqa, Grohs:2022fyq, Froustey:2023skf, Grohs:2023pgq, Nagakura:2025hss,Fiorillo:2025gkw}, investigating both the conditions under which FFIs can develop and approaches to modeling them in classical simulations using quantum moment methods~\cite{Strack:2005ux,Myers:2021hnp,Kneller:2024buy,Froustey:2024sgz,Grohs:2025ajr}. So far, theoretical studies have shown that flavor instabilities driven by neutrino self-interactions can lead to rapid flavor conversion on nanosecond timescales—far shorter than hydrodynamic timescales—making them nearly impossible to model in dynamical simulations, though in-situ flavor transformation could prevent the development of instabilities with timescales as short as those predicted by post-processing models using classical transport \cite{Nagakura:2022kic, Nagakura:2022xwe, Fiorillo:2024qbl,Liu:2024nku,Fiorillo:2024uki,Xiong:2024tac,Johns:2025mlm}.

Despite the challenges of modeling flavor conversions in global simulations, several studies have explored their potential impact in post-merger accretion disks~\cite{Li:2021vqj, Just:2022flt, Fernandez:2022yyv, Mukhopadhyay:2024zzl}, including a recent work~\cite{Lund:2025jjo} that incorporates fast flavor conversions in an angle-dependent neutrino transport scheme. These works have found that, depending on the lifetime of the remnant neutron star~\cite{Fernandez:2022yyv}, neutrino flavor conversions can make the disk winds more neutron rich, thus enhancing the r-process in the outflow. However, they focus on a phase of the evolution that does not encompass the highly dynamical merger phase—when the structure and composition of the disk are initially set~\cite{Camilletti:2024otr}. In-situ parameterized flavor transformation has also been realized in the context of core-collapse supernovae, with results suggesting its important impacts on dynamics and nucleosynthesis~\cite{Xiong:2020ntn, Ehring:2023lcd, Ehring:2023abs,Xiong:2024pue, Wang:2025nii, Akaho:2025giw, Padilla-Gay:2025tko, Bhattacharyya:2025gds, Xiong:2025fge,Cornelius:2025tyt}. Furthermore, there is growing interest in comparing the occurrence and growth rates of FFI and other relevant mechanisms~\cite{Padilla-Gay:2025tko}, such as collisional flavor instabilities~\cite{Nagakura:2025hss, Wang:2025vbx}. Overall, the exact neutrino flavor evolution triggered by a combination of various possible instabilities with different characteristic times remains elusive. Efforts need to be made in both dynamical timescale with in-situ simulations and sub-nano second timescale with quantum kinetics to understand more quantitatively the outcome of the neutrino flavor conversions in dense environments.

Our previous work~\cite{Qiu:2025kgy} considered neutrino flavor conversions in dynamical-spacetime BNS merger simulations. We employed a Bhatnagar-Gross-Krook (BGK) subgrid operator, using approximations of the flavor instability growth rates and assumptions on the flavor equilibrium asymptotic states. Our findings suggest that when neutrino flavor conversions are enabled, electron (anti)neutrinos are converted to heavy-lepton types, i.e., muon- and tau-type neutrinos. Through charged current weak interactions, their impacts on the composition and structure of the remnant manifest as the ejecta become more neutron rich, potentially leaving imprints on the post-merger GW signals. 

In this study, we extend our previous analyses by considering two equations of state (EoS), as well as different mixing prescriptions. For the considered binary configuration, the different EoS leads to the formation of a massive remnant or of a black hole, allowing us to explore two different post-merger scenarios.
Moving beyond the density-dependent flavor mixing approach, we also implement, for the first time in BNS merger simulations, the neutrino flavor conversions induced by FFI, based on electron lepton number (ELN) flux factor criterion~\cite{Abbar:2020fcl,Richers:2022zug}. Furthermore, we vary the flavor conversion relaxation times, conduct analysis on angular dependence of the ejecta and flavor composition evolutions and study the correlation of the neutrino and GW emissions. Our results also reinforce our previous finding that flavor conversions can significantly alter nucleosynthetic yields from BNS mergers. Some isobaric abundances change by up to an order of magnitude, highlighting the importance of including flavor physics in future simulations.

\section{Methods}

\begin{figure*}
\includegraphics[width=1.6\columnwidth]{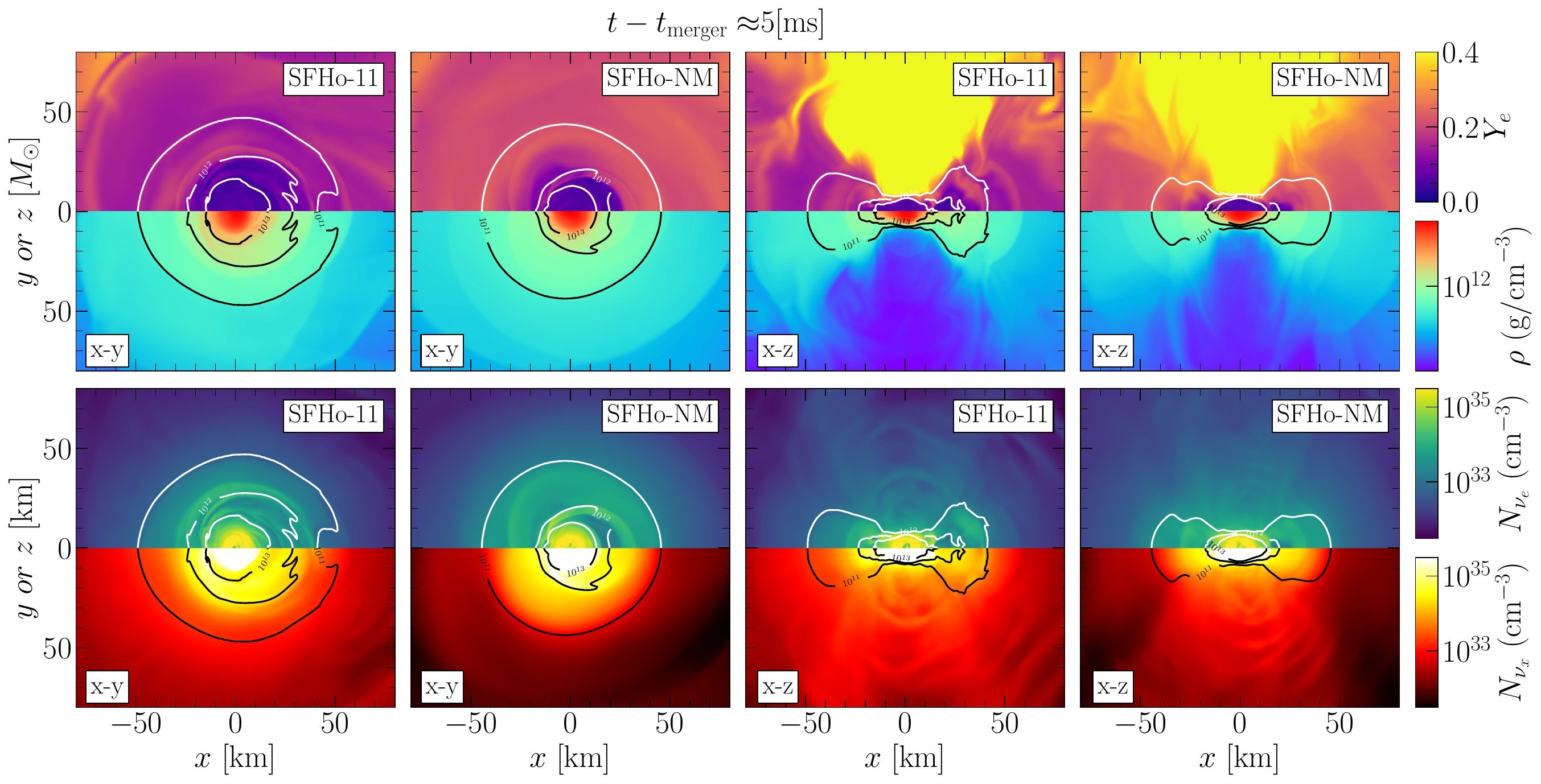}
\includegraphics[width=1.6\columnwidth]{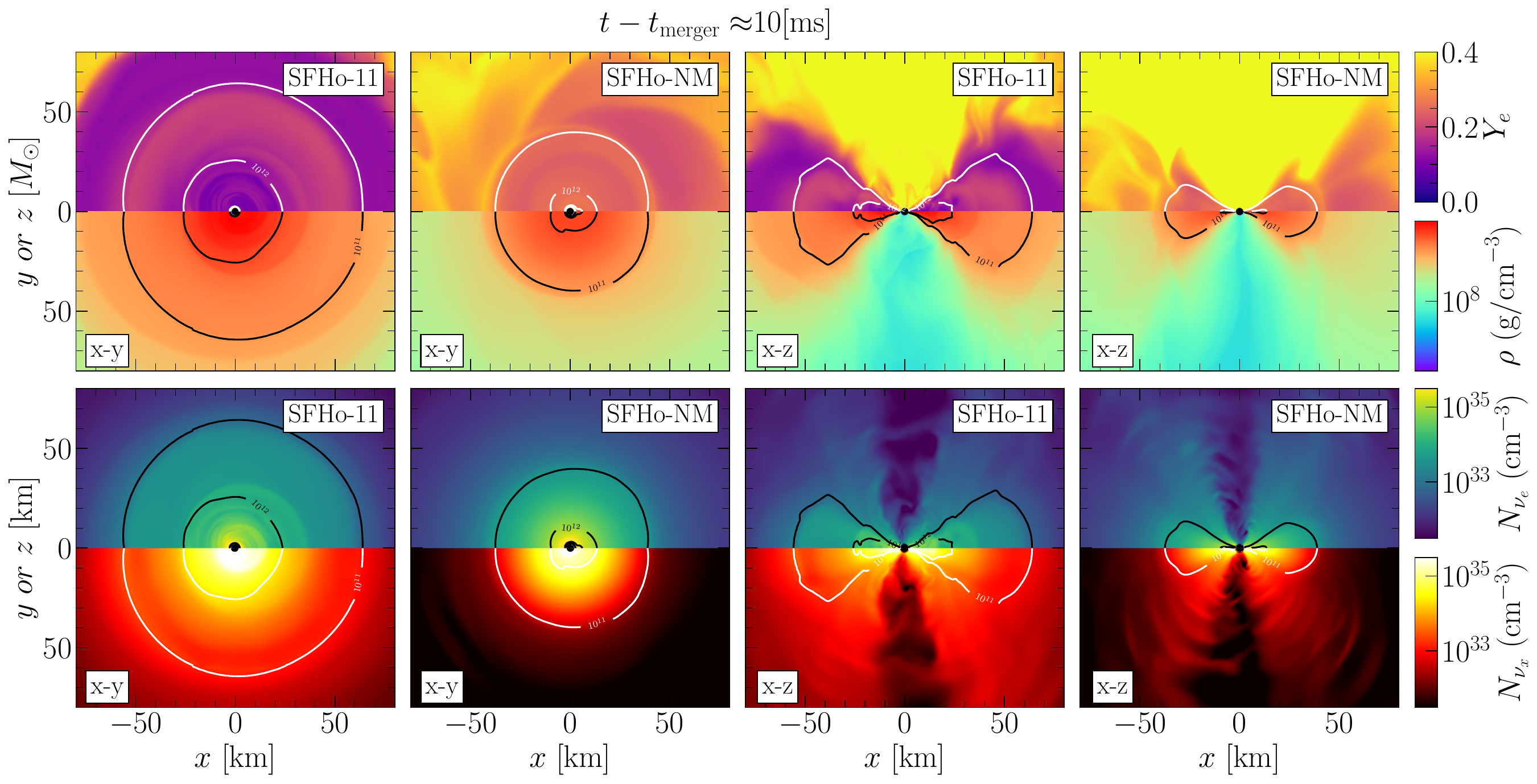}
\includegraphics[width=1.6\columnwidth]{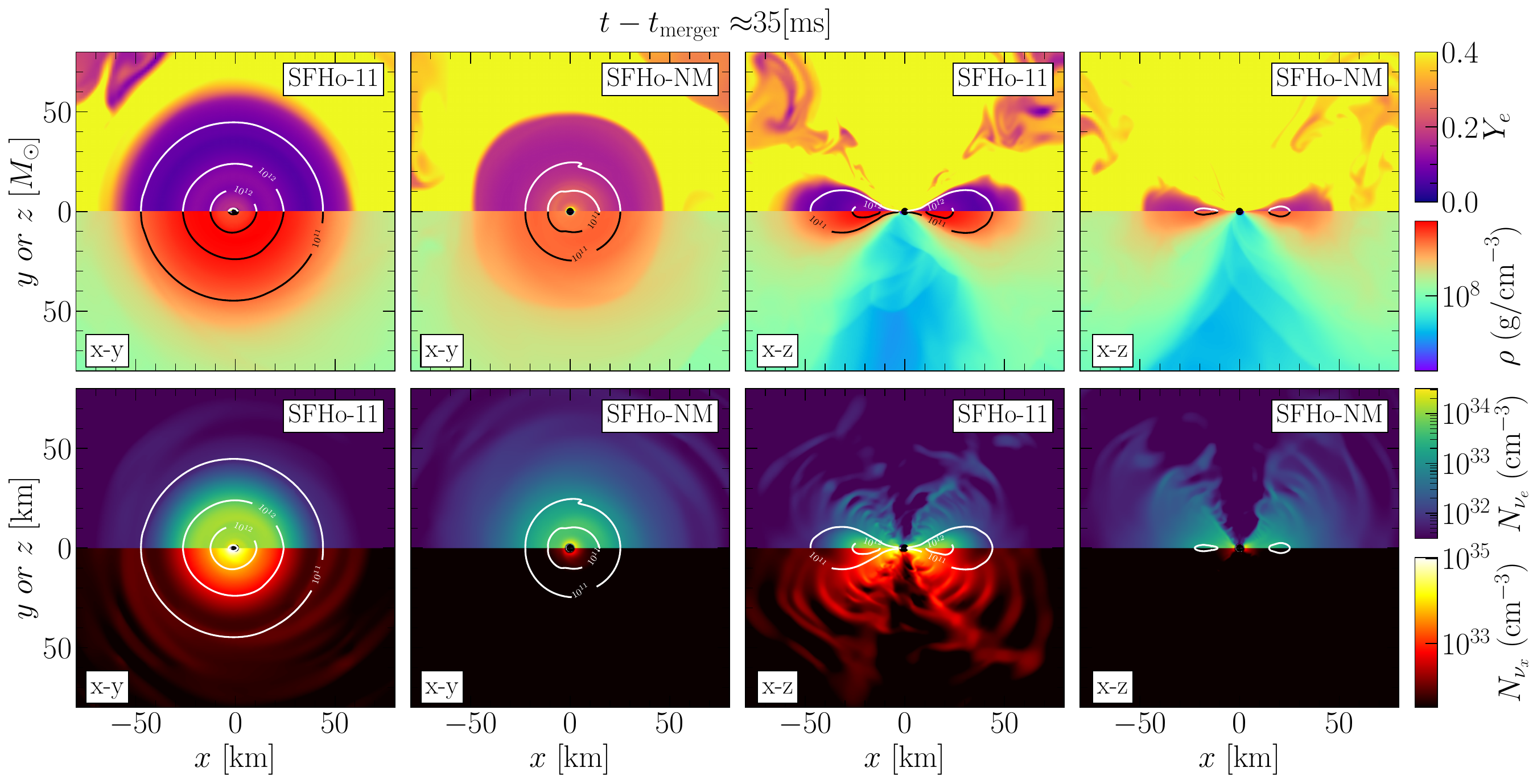}
\caption{Simulation snapshots in the equatorial (xy) and meridional (xz) planes for the SFHo-11 and the SFHo-NM models at $\sim$5, 10 and 35 ms after merger on the left (right) four panels, respectively. The top-row panels of each plot show electron fractions (upper half) and rest mass densities (lower half). The second-row panels of each plot display electron neutrino number densities (upper half) and heavy-lepton neutrino number densities (lower half). The contour lines denote where the density is at $10^{11}$, $10^{12}$ and $10^{13}$$\mathrm{g/cm}^3$. The $\sim$5 ms post-merger snapshots correspond to dynamics just before collapse to BHs, while in $\sim$10 ms post-merger snapshots the collapses have happened. The region with lapse function $\alpha \leq 0.15$ (black circles) denotes the approximate location of the BH apparent horizon. Finally, the $\sim$35 ms post-merger snapshots depicts the states of the system at which the remnant BHs enter the accretion phases. Due to the uncertainties of the collapse times, SFHo-11 and SFHo-NM simulations appear to have very different patterns of density distributions after the collapses, which are not due to the neutrino flavor conversion effects. In terms of the composition of the disk and ejecta, we find similar results as in~\cite{Qiu:2025kgy}. In different stages of the remnant evolutions, the disk in the SFHo-11 mixing model is more neutron rich than that of the SFHo-NM model. Due to flavor conversion effects transforming electron (anti-)neutrinos to heavy-lepton neutrinos, we also find that electron neutrinos are more abundant in the SFHo-NM than those of SFHo-11, and vice versa for heavy-lepton neutrinos.}
\label{fig:2d}
\end{figure*}

\subsection{Simulation setup}
Our simulations are based on \texttt{THC\_M1}~\cite{Radice:2012cu, Radice:2013hxh, Radice:2013xpa, Radice:2021jtw}, a general relativistic hydrodynamics code, incorporating truncated moment neutrino transport mechanisms with BGK operator for neutrino flavor oscillations. The M1 scheme and closure treatments are consistent with those in~\cite{Qiu:2025kgy}, in which we consider the (neutrino) radiation stress energy tensors of each neutrino flavor as
\begin{equation*}
T_{\mathrm{rad}}^{\alpha \beta}=E n^\alpha n^\beta+F^\alpha n^\beta+n^\alpha F^\beta+P^{\alpha \beta}\,\,,
\end{equation*}
where $n^\alpha$ is the unit vector normal to the constant time hypersurfaces, such that  $E$ is the neutrino radiation energy density, $F^\alpha$ is the neutrino radiation flux, and $P^{\alpha \beta}$ is the radiation pressure tensor. These are the zeroth, first, and second order angular moments, respectively. The essence of the M1 scheme is to relate $P^{\alpha \beta}$ with $E$ and $F^\alpha$, for which we adopt a standard closure equation~\cite{Shibata:2011kx,Foucart:2015vpa}, as discussed in details in Supplemental Material of~\cite{Qiu:2025kgy}. Apart from the energy and flux density, we consider the neutrino number density of each flavor in the fluid frame
\begin{equation}
n=-N^\alpha u_\alpha \, ,
\end{equation}
where $u^\alpha$ is the fluid 4-velocity and $N^\alpha$ is the neutrino number current density  (see more discussions in~\cite{Radice:2021jtw}) defined as
\begin{equation}
N^\alpha = n f^\alpha = n\left(u^\alpha + \frac{H^\alpha}{J}\right).
\end{equation}
Here $J$ is the neutrino energy density measured in the fluid rest frame, $H^\alpha$ is the neutrino momentum density in the fluid frame, and $f^\alpha$ is the dimensionless 4-vector specifying the mean neutrino propagation direction. The spatial components of $f^\alpha$ are
\begin{equation}
f^i = W\left(v^i - \frac{\beta^i}{\alpha}\right) + \frac{H^i}{J},
\end{equation}
where $\alpha$ is the lapse function, $\beta^i$ the shift vector, $W$ is the fluid Lorentz factor, $v^\alpha= \gamma^\alpha{ }_\beta u^\beta \ W$ is the fluid three velocity, and $\gamma^\alpha{ }_\beta$ projects vectors onto the hypersurface normal to $n^\alpha$. The time component of $f^\alpha$ follows
\begin{equation}
\Gamma=\alpha f^0=W-\frac{1}{J} H^\alpha n_\alpha \, ,
\end{equation}
in which $\Gamma$ essentially plays the role of a generalized Lorentz factor for the neutrino number density in the fluid frame.

The M1 equations can be outlined as
\begin{equation}
    \partial_t \boldsymbol{U} + \partial_i \boldsymbol{F}^i(\boldsymbol{U}) = \boldsymbol{G}(\boldsymbol{U}) + \boldsymbol{S}(\boldsymbol{U})
\end{equation}
where the evolved variables are
\begin{equation}
\boldsymbol{U}=\left(\begin{array}{c}
\sqrt{\gamma} n \Gamma \\
\sqrt{\gamma} E \\
\sqrt{\gamma} F_k
\end{array}\right).
\end{equation}
Here, $\gamma = \det \gamma_{ij}$ is the determinant of the spatial metric. The flux terms are
\begin{equation}
  \boldsymbol{F}^i = \left( \begin{array}{c}
\alpha \sqrt{\gamma} n f^i \\
\sqrt{\gamma}\left[\alpha F^i - \beta^i E\right] \\
\sqrt{\gamma}\left[\alpha P^i{ }_k - \beta^i F_k\right]
\end{array} \right) \, .
\end{equation}
The source terms are
\begin{equation}
\begin{aligned}
&\boldsymbol{S}=\left(\begin{array}{c}
\alpha \sqrt{\gamma}\left[\eta^0-\kappa_a^0 n\right] \\
-\alpha \sqrt{\gamma} \mathcal{S}^\mu n_\mu \\
\alpha \sqrt{\gamma} \mathcal{S}^\mu \gamma_{k \mu}
\end{array}\right),
\end{aligned}
\end{equation}
where $\eta^0$ is the neutrino number emission coefficient, $\kappa_a^0$ the neutrino number absorption coefficient, and $\mathcal{S}^\mu$ the interaction four-vector responsible for energy and momentum sources
\begin{equation}
\mathcal{S}^\mu=\left(\eta-\kappa_a J\right) u^\mu-\left(\kappa_a+\kappa_s\right) H^\mu,
\end{equation}
where $\eta, \kappa_a$, and $\kappa_s$ are properly energy integrated neutrino emissivity, absorption, and scattering coefficients. Note that all the opacities are calculated in the fluid frame, with a weak interaction calculation code considering all relevant reactions shown in Table. 1 of Supplemental Materials in~\cite{Qiu:2025kgy}. Also note that the term $\boldsymbol{S}$, describing the coupling with matter, is usually stiff and has to be solved implicitly with a root-finding process. Finally, the geometric source terms are
\begin{equation}
\begin{aligned}
&\boldsymbol{G}=\left(\begin{array}{c}
0 \\
\alpha \sqrt{\gamma}\left[P^{i k} K_{i k}-F^i \partial_i \log \alpha\right] \\
\sqrt{\gamma}\left[F_i \partial_k \beta^i-E \partial_k \alpha+\frac{\alpha}{2} P^{i j} \partial_k \gamma_{j i}\right]
\end{array}\right) \, ,
\end{aligned}
\end{equation}
where $K_{ik}$ is the extrinsic curvature. The above equations are solved after the spacetime and hydrodynamics evolution in each iteration of the simulations. 

Our initial configurations consist of two equal-mass binaries of non-rotating NSs, each with a gravitational mass of $1.35\,M_{\odot}$. The stars are placed at an initial separation of $45\,\mathrm{km}$. These initial data are generated using the \texttt{Lorene} pseudo-spectral code~\cite{Gourgoulhon:2000nn}. We employ two EoSs for the two considered binary systems: the DD2 EoS~\cite{Typel:2009sy,Hempel:2009mc} and the SFHo EoS~\cite{Steiner:2012xt}. Adaptive mesh refinement (AMR)~\cite{Berger:1984zza,Berger:1989a} is used by employing the \texttt{Carpet} driver~\cite{Schnetter:2003rb, Reisswig:2012nc} of the \texttt{Einstein Toolkit}~\cite{Loffler:2011ay,roland_haas_2024_14193969}, with 7 refinement levels. The innermost refinement region tracks movements of each neutron star prior to merger and later resolves the central region of the remnant. We consider two resolutions, low resolution (LR) and standard resolution (SR), with corresponding finest grid spacings $0.167\,M_{\odot} \simeq 246\,\mathrm{m}$ and $0.125\,M_{\odot} \simeq 184\,\mathrm{m}$.

\begin{table*}
\centering
\caption{Overview of all simulation setups and their key diagnostics, including the names of simulation, resolutions, EoS types, neutrino flavor mixing threshold densities $\rho_\mathrm{mixing}$, mixing relaxation times, mixing prescriptions, post-merger time to collapse to BHs, masses of total ejecta, masses of neutron-rich ($Y_e<0.25$) ejecta and ejecta velocities at infinity. Note that the ejecta is measured on a sphere located at $\sim 295$ km away from the center of simulation domain, based on geodesic criterion.}
\begin{tabular}{l c c c c c c c c c c}
\toprule
\hline
Simulation & Resolution & EoS & \makecell{$\rho_\mathrm{mixing}$ \\ $[\mathrm{g/cm}^3]$} & $\tau_a$ [ns] & Prescription & \makecell{$t_{\mathrm{coll}}-t_{\mathrm{merg}}$ \\ $\,[\mathrm{ms}]$ } & \makecell{$M^{\mathrm{ej}}_\mathrm{total}$ \\ $[10^{-2} M_{\odot}]$} & \makecell{$M^{\mathrm{ej}}_{Y_e<0.25}/M^{\mathrm{ej}}_\mathrm{total}$ \\ $[\%]$} & \makecell{$M^{\mathrm{ej}}_{Y_e<0.15}/M^{\mathrm{ej}}_\mathrm{total}$ \\ $[\%]$} & $v_{\infty}$ [$c$] \\
\midrule
\hline
DD2-NM       & LR & DD2  & --        & --   & --    & --     & 0.1730 & 20.88 & 10.22 & 0.1530 \\
DD2-11       & LR & DD2  & $10^{11}$ & 0.5  & MB  & --     & 0.1480 & 59.45 & 28.29 & 0.1704 \\
DD2-13       & LR & DD2  & $10^{13}$ & 0.5  & MB  & --     & 0.08940 & 43.38 & 22.99 & 0.1673 \\
DD2-FFI      & LR & DD2  & $10^{13}$ & 0.5  & MX      & --     & 0.08761 & 25.84 & 17.60 & 0.1617 \\
DD2-13-slow  & LR & DD2  & $10^{13}$ & 50.0 & MB  & --     & 0.1351 & 52.32 & 31.36 & 0.1714 \\
DD2-NM       & SR & DD2  & --        & --   & --    & --     & 0.2096 & 22.40 & 10.44 & 0.1657 \\
DD2-11       & SR & DD2  & $10^{11}$ & 0.5  & MB  & --     & 0.1676 & 71.19 & 39.50 & 0.1608 \\
DD2-13       & SR & DD2  & $10^{13}$ & 0.5  & MB  & --     & 0.1312 & 43.22 & 23.40 & 0.1631 \\
DD2-FFI      & SR & DD2  & $10^{13}$ & 0.5  & MX      & --     & 0.1077 & 43.23 & 30.71 & 0.1888 \\
SFHo-NM      & LR & SFHo & --        & --   & --    & 10.26  & 0.6900 & 39.51 & 10.19 & 0.2267 \\
SFHo-11      & LR & SFHo & $10^{11}$ & 0.5  & MB  & 8.215  & 0.5420 & 77.85 & 40.82 & 0.2082 \\
SFHo-13      & LR & SFHo & $10^{13}$ & 0.5  & MB  & 14.55  & 0.4517 & 69.62 & 39.27 & 0.2189 \\
SFHo-FFI     & LR & SFHo & $10^{13}$ & 0.5  & MX      & 17.27  & 0.5387 & 38.71 & 8.703 & 0.2345 \\
SFHo-NM      & SR & SFHo & --        & --   & --    & 8.078  & 0.5889 & 46.72 & 17.52 & 0.2102 \\
SFHo-11      & SR & SFHo & $10^{11}$ & 0.5  & MB  & 11.15  & 0.4244 & 67.95 & 32.14 & 0.2239 \\
SFHo-13      & SR & SFHo & $10^{13}$ & 0.5  & MB  & 5.513  & 0.5252 & 74.89 & 38.70 & 0.2517 \\
SFHo-FFI     & SR & SFHo & $10^{13}$ & 0.5  & MX      & 5.607  & 0.6885 & 65.76 & 25.93 & 0.1829 \\
\bottomrule
\hline
\end{tabular}
\label{tab:setup}
\end{table*}

\begin{figure*}
\includegraphics[width=2.0\columnwidth]{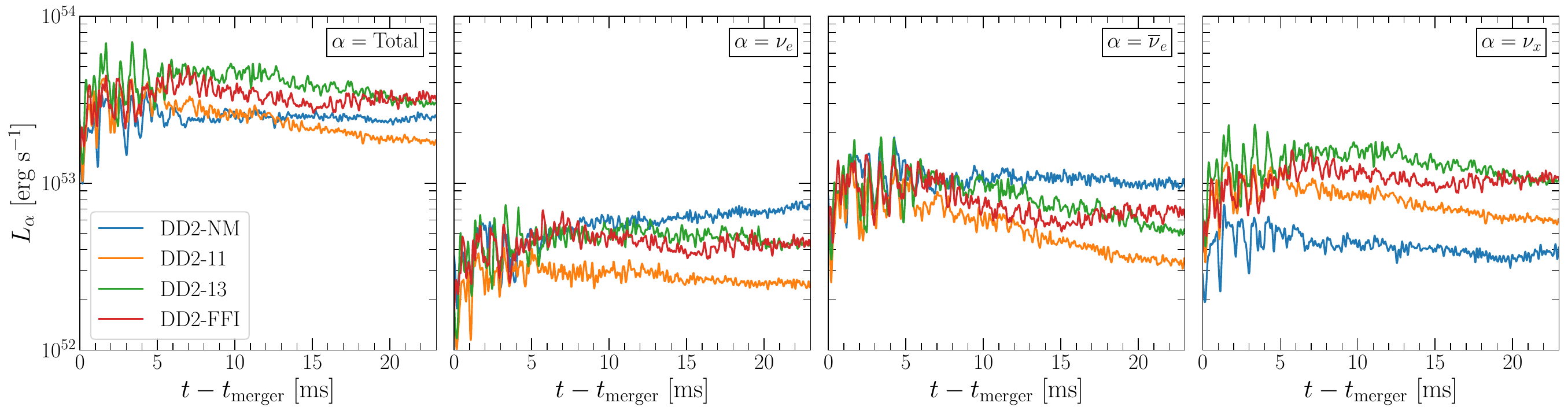}
\includegraphics[width=2.0\columnwidth]{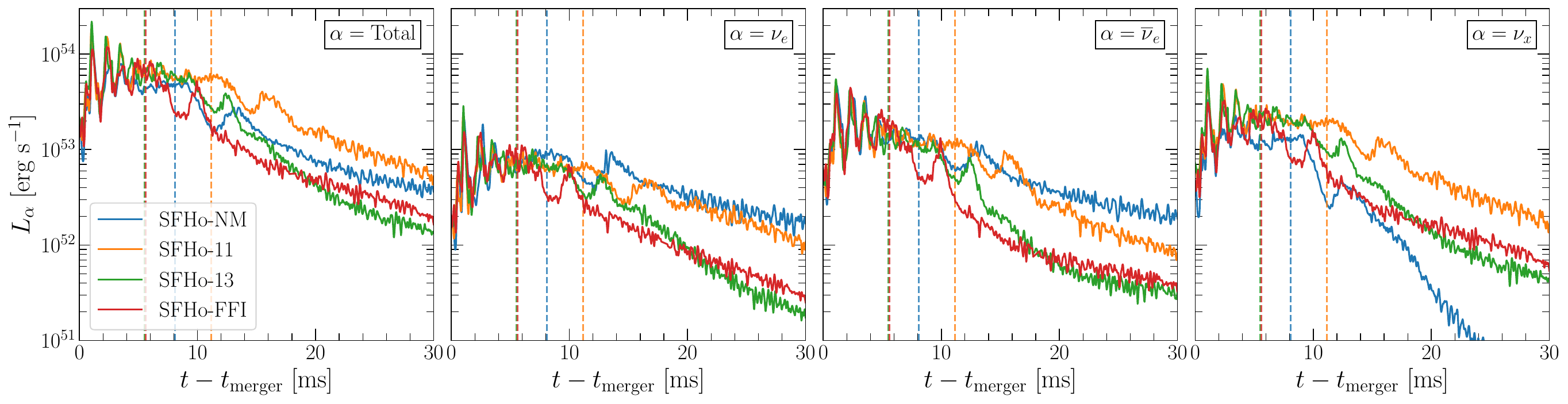}
        \caption{Post-merger total neutrino luminosity, and luminosity for $\nu_e$, $\bar{\nu}_e$, and $\nu_x$ ($= \nu_\mu + \nu_\tau$), respectively, from left to right panels in each row. Note that $\nu_x$ luminosity does not contain the contributions from heavy-lepton anti-neutrinos. The upper row shows results of simulations considering DD2 EoS and the lower row shows those of SFHo EoS. We mark the collapse times for the SFHo runs with dashed lines. Neutrino flavor conversions $\nu_e, \bar{\nu}_e \rightarrow \nu_x, \bar{\nu}_x$ generically increase the $\nu_x$ luminosities, while decreasing the electron (anti-)neutrino luminosities. Among the three mixing models with DD2 EoS, the DD2-13 simulation has the highest net luminosities compared to the DD2-11 and the DD2-FFI simulations. This is because it includes additional flavor conversion effects in the inner disk regions between density of $10^{11}$ and $10^{13}\mathrm{g/cm}^3$ compared to the DD2-11, and does not filter out the FFI-stable regions as compared to the DD2-FFI. For the SFHo runs, we do not observe a clear hierarchy, due to differences in the collapse times and remnant disk masses. At late times, the SFHo-11 has higher luminosities for $\nu_x$ while the other mixing models are generally comparable. We note that such differences are likely not results of the neutrino flavor conversion effects, though.}
        \label{fig:lumi}
\end{figure*}

\subsection{Neutrino flavor conversion prescriptions}
Neutrino oscillation effects manifest over different timescales in different environments. In particular, neutrino self-interaction results in fast flavor oscillations in neutrino dense environments such as BNS mergers. However, evolution on sub-nano second time scales are required to tackle that explicitly.
Instead of resolving the explicit form of the mean field quantum kinetic equations for the density matrices of each flavor of neutrino fields, our BGK model uses a collisional operator~\cite{Bhatnagar:1954zz, Nagakura:2023jfi, Qiu:2025kgy} to parameterize the flavor conversion
\begin{equation}
S_{\mathrm{FC}}=\frac{1}{\tau_a}\left(f^\text{eq}-f\right),
\end{equation}
where $\tau_a$ is the relaxation time to reach a given asymptotic equilibrium state, $f^\text{eq}$, starting from an arbitrary neutrino distribution function, $f$. \reply{In particular, the BGK subgrid model provides a more accurate treatment than the effective, instantaneous flavor equilibration prescriptions used in, e.g.,~\cite{Li:2021vqj,Just:2022flt}, since those methods would introduce errors by approximating continuous evolution as a sequence
of discrete instabilities~\cite{Urquilla:2025idk}.} 

We apply an operator-splitting approach to incorporate the BGK flavor conversion term, $S_{\mathrm{FC}}$, in our M1 scheme. In the flavor mixing relaxation sub-step, taking place after the source term updates and the feed back to the fluid, the neutrino densities, energies and fluxes are updated according to
\begin{equation}
\begin{aligned}
&n^{\text{new}} = \lambda\, n^{\text{old}} + (1 - \lambda)\, n^{\text{eq}},\\
&E^{\text{new}} = \lambda\, E^{\text{old}} + (1 - \lambda)\, E^{\text{eq}},\\
&F_i^{\text{new}} = \lambda\, F_i^{\text{old}} + (1 - \lambda)\, F_i^{\text{eq}},
\end{aligned}
\label{eq:lambda}
\end{equation}
where the superscripts \(\text{old}\) and \(\text{new}\) denote the values before and after this BGK relaxation step, respectively. Here $\lambda = e^{-\Delta t / \tau_a}$, in which $\Delta t$ is the same as the M1 time step. We note that the updates are conducted on neutrino number density and two moments in the lab frame. It is useful to highlight the global $\mathrm{SU}(3)$ symmetry in flavor space the neutrino self-interaction Hamiltonian possesses, which leads to an overall conservation of lepton number in the absence of collisions and external sources. This places constraints on $f^\text{eq}$, and our choices must be compatible with this symmetry.

As aforementioned, the BGK operator requires inputs of relaxation times and asymptotic states, for which we now introduce two treatments. The first treatment is the same as that in our previous study~\cite{Qiu:2025kgy}. This treatment assumes a piecewise-constant, density dependent relaxation time, $\tau_a$, such that $\tau_a=0.5$ ns for $\rho < \rho_{\mathrm{mixing}}$, and $\tau_a=\infty$ otherwise. We notice that infinite relaxation times corresponds to the absence of flavor conversions. \reply{The choice of $\tau_a=0.5$ ns is motivated by the fact that it is comparable to the typical flavor conversion timescales reported in local linear stability analysis of fast flavor conversions in neutron star mergers~\cite{Froustey:2023skf}.} The corresponding asymptotic states, which are motivated by quantum many-body calculations and possible beyond standard model effects~\cite{Martin:2023gbo}, are noted as `MB' (short for many-body). \reply{MB assumes detailed balance among the pairwise reactions, such that the rate—which depends on the product of the reactant densities—for converting one flavor pair into another flavor pair equals the rate for the reversed reaction, so}
\begin{equation}
n_e^\text{eq} \bar{n}_e^\text{eq}=\frac{1}{4}n_x^\text{eq} \bar{n}_x^\text{eq} \, ,
\end{equation}
where $n_e$ is the number density of electron neutrinos, and $n_x$ is the number density of heavy-lepton neutrinos (muon and tau neutrinos combined, under the 4 neutrino species framework we consider in this work), with their anti-particles denoted under bars. Note that such asymptotic states conserve the ELN and total lepton numbers, with the conservation enforced for the following 
\begin{equation}
\begin{aligned}
& N=n_e+n_x+\bar{n}_e+\bar{n}_x \\
& N_e=n_e-\bar{n}_e \\
& N_x=n_x-\bar{n}_x
\end{aligned}    
\end{equation}
Based on the mixing equilibrium found for the number density, we then calculate a $4\times 4$ mixing matrix $Y_{a b}$ as 
\begin{equation}
\begin{aligned}
Y_{a,a} &= \min\left(1, \frac{n_a^\text{eq}}{n_a}\right) \\
Y_{2,0} &= 1 - Y_{0,0}, & \text{(}\nu_e \rightarrow \nu_x\text{)} \\
Y_{3,1} &= 1 - Y_{1,1}, & \text{(}\bar{\nu}_e \rightarrow \bar{\nu}_x\text{)} \\
Y_{0,2} &= 1 - Y_{2,2}, & \text{(}\nu_x \rightarrow \nu_e\text{)} \\
Y_{1,3} &= 1 - Y_{3,3}, & \text{(}\bar{\nu}_x \rightarrow \bar{\nu}_e\text{)} \\
Y_{a,b} &= 0, \quad \text{for all other } (a,b).
\end{aligned}
\end{equation}
Such $Y_{a b}$ represents the flavor conversion kernel, 
\begin{equation}
n_a^{\text{eq}}=Y_{a b} n_b
\end{equation}
while respecting the conservation laws we enforce. Then we mix the energy and fluxes (for different directions separately) as
\begin{equation}
E_a^{\text{eq}}=Y_{a b} E_b, \quad  F_a^{\text{eq}}=Y_{a b} F_b.
\end{equation}

The second treatment we consider is informed by the fast flavor conversion analysis. Below a density threshold  $\rho_{\mathrm{mixing}}$, we identify regions susceptible to fast instabilities based on ELN flux factor~\cite{Abbar:2020fcl,Richers:2022dqa}
\begin{equation}
f_\mathrm{ELN}=\frac{\frac{F_{\nu_e}}{E_{\nu_e}} N_{\nu_e}-\frac{F_{\bar{\nu}_e}}{E_{\bar{\nu}_e}} N_{\bar{\nu}_e}}{N_{\nu_e}-N_{\bar{\nu}_e}} \, .
\label{eq:ELN_flux}
\end{equation}
When the ELN flux factor is larger or equal to 1, we identify the corresponding cell to be unstable to fast flavor conversion since it requires necessarily that there is a crossing in the ELN distribution (although it does not flag every unstable location). Therefore, we apply the BGK subgrid model to relax the neutrino flavor to its asymptotic equilibrium state with a given relaxation time, which is also set to be a constant $\tau_a=0.5$~ns. The corresponding asymptotic equilibrium states are noted as `MX' (short for maximal). The MX mixing prescription used for the density dependent mixing model gives
\begin{equation}
\begin{aligned}
& n_e^\text{eq}=N / 6+N_e / 2 \, , \\
& \bar{n}_e^\text{eq}=N / 6-N_e / 2 \, , \\
& n_x^\text{eq}=N / 3+N_x / 2 \, , \\
& \bar{n}_x^\text{eq}=N / 3-N_x / 2 \, ,
\end{aligned}
\end{equation}
which can be solved analytically. Identical procedures then take place to compute the $Y_{a b}$ and then obtain the energy and fluxes mixing quantities.

In summary, we consider two neutrino flavor conversion prescriptions for the mixing equilibrium states, namely MB and MX. We run in total 17 simulations with various setups, as shown in Table.~\ref{tab:setup}. We consider different density thresholds $10^{11}$$\mathrm{g/cm}^3$ and $10^{13}$$\mathrm{g/cm}^3$, below which to turn on the flavor conversion effects. As $\rho_{\mathrm{mixing}}$ we explore $10^{11}$$\mathrm{g/cm}^3$ (approximately where the electron (anti)neutrinos decouple) and $10^{13}$$\mathrm{g/cm}^3$ (approximately where the heavy-lepton neutrinos decouple the transition between the central remnant and the disk occurs). For the FFI models, we not only apply the density cutoff  $\rho_{\mathrm{mixing}}=10^{13}$$\mathrm{g/cm}^3$, but also the ELN flux factor criterion in Eq.~\ref{eq:ELN_flux}, to activate the flavor conversions. These mixing models are then compared to the classical runs, i.e., no-mixing neutrino M1 simulations, denoted as DD2(SFHo)-NM. All models are run at LR and SR. We find quantitative, but not qualitative, differences between the resolutions. When not otherwise stated, we report data from the SR resolutions in the discussion below. We limit our discussion to those features that are robust across resolutions.

\begin{figure*}
\includegraphics[width=2.0\columnwidth]{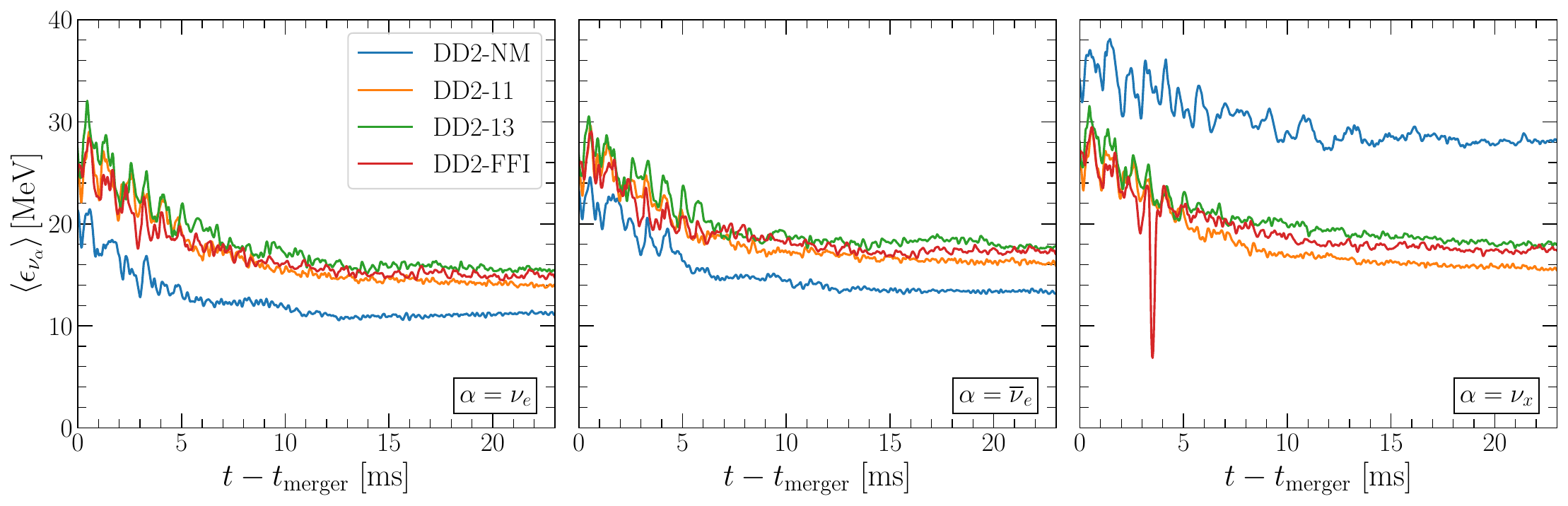}
\includegraphics[width=2.0\columnwidth]{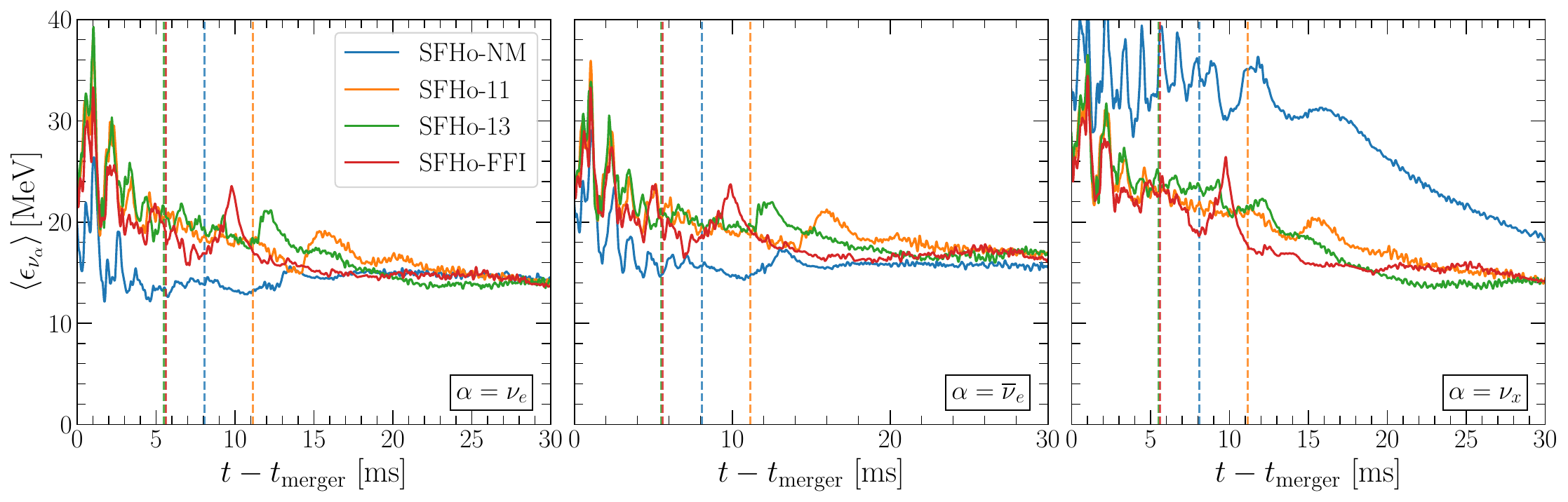}
        \caption{Post-merger neutrino mean energy for $\nu_e$, $\bar{\nu}_e$, and $\nu_x$, respectively, from left to right panels in each row. The upper row shows results of simulations considering DD2 EoS and the lower row shows those of SFHo EoS. We mark the collapse times for the SFHo runs with dashed lines. For the no-mixing models, heavy-lepton neutrinos generally have higher mean energies, indicating their smaller decoupling radii compared to the electron (anti)neutrinos. For the neutrino mixing models DD2(SFHo)-11, DD2(SFHo)-13 and DD2(SFHo)-FFI, the mean energies of all flavors are comparable, because flavor conversions can exchange energy of different neutrinos. The mean energy of $\nu_e$ seems a bit smaller, but it could be due to the fact that high energy $\nu_e$ are absorbed more easily by the neutron rich material and the spectrum tends to become a bit softer. The mixing models mostly have higher $\nu_e$ and $\bar{\nu}_e$ mean energies and lower $\nu_x$ mean energy compared to the no-mixing model. \reply{Note that the sharp dip in the DD2-FFI model at around $\sim$3.5 ms post-merger is caused by a transient outburst in the $\nu_x$ number flux. This outburst is generated when a parcel of material crosses a refinement-level corner in the mesh, producing a short–lived numerical artifact rather than a physical signal.}}
        \label{fig:mean_energy}
\end{figure*}

\section{Results}

\begin{figure*}
\includegraphics[width=2.0\columnwidth]{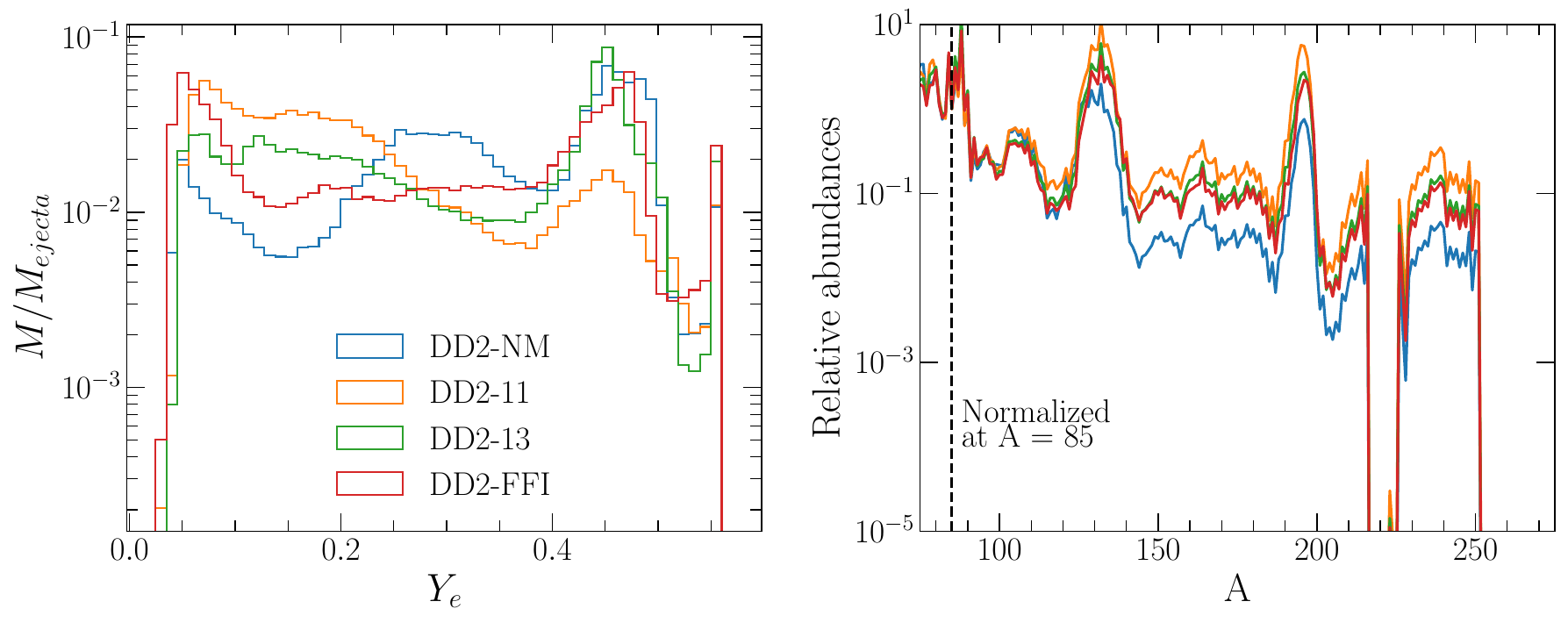}
\includegraphics[width=2.0\columnwidth]{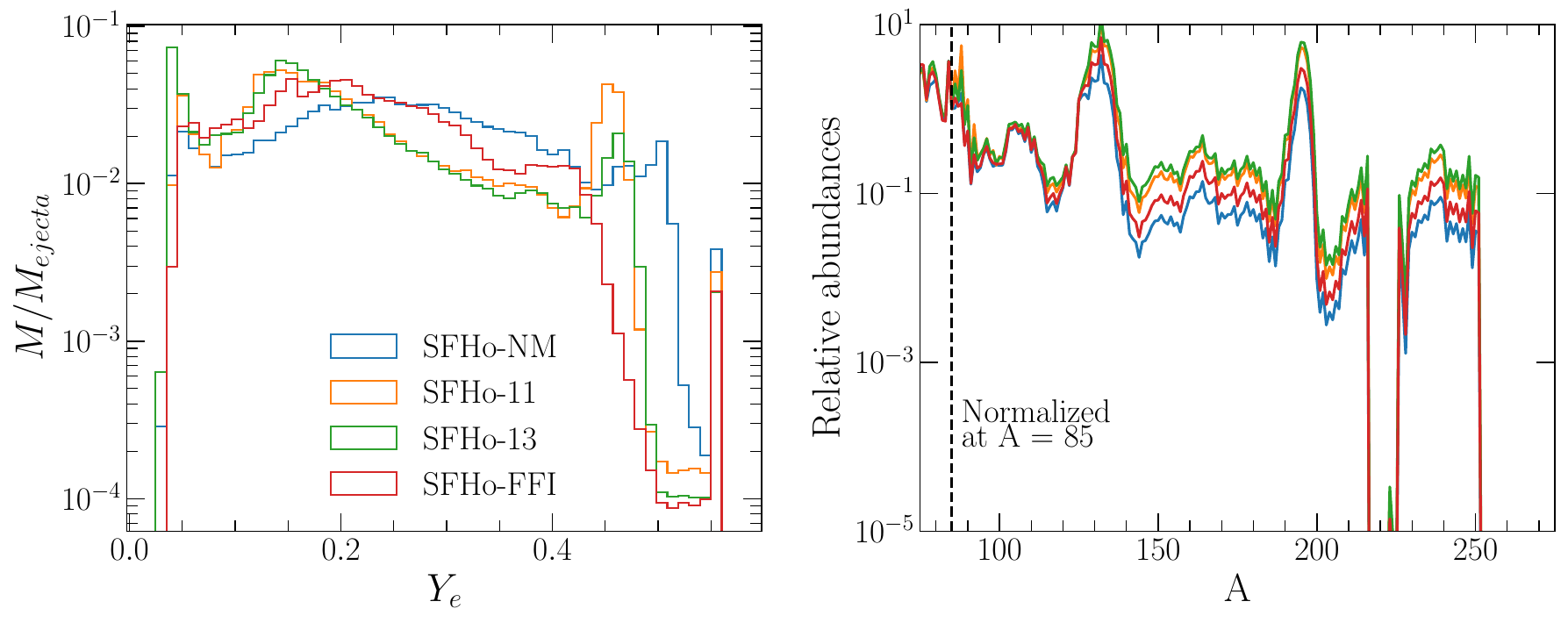}
    \caption{Upper (lower) left panels: histograms of the electron fraction distributions of the ejecta for the four DD2 (SFHo) models. Upper (lower) right panel: the relative abundances of nuclei of mass number $A$ formed in the ejecta of four DD2 (SFHo) models. The relative neutron richness of the ejecta follows the order: DD2-11, DD2-13, DD2-FFI, and DD2-NM. Notably, the ordering between the DD2(SFHo)-11 and DD2(SFHo)-13 is reversed for the two different EoS's. On the right plots, we normalize the yields at $A=85$. The yields relative to $A=85$ for lanthanides and heavy elements are one order of magnitude higher in the DD2-11 simulation than the DD2-NM simulation. Such difference shrinks in the SFHo simulations, but is still up to a factor of five between the SFHo-11 and SFHo-NM heavy element yields.} 
    \label{fig:ejecta}
\end{figure*}

\begin{figure}
\includegraphics[width=0.98\columnwidth,keepaspectratio]{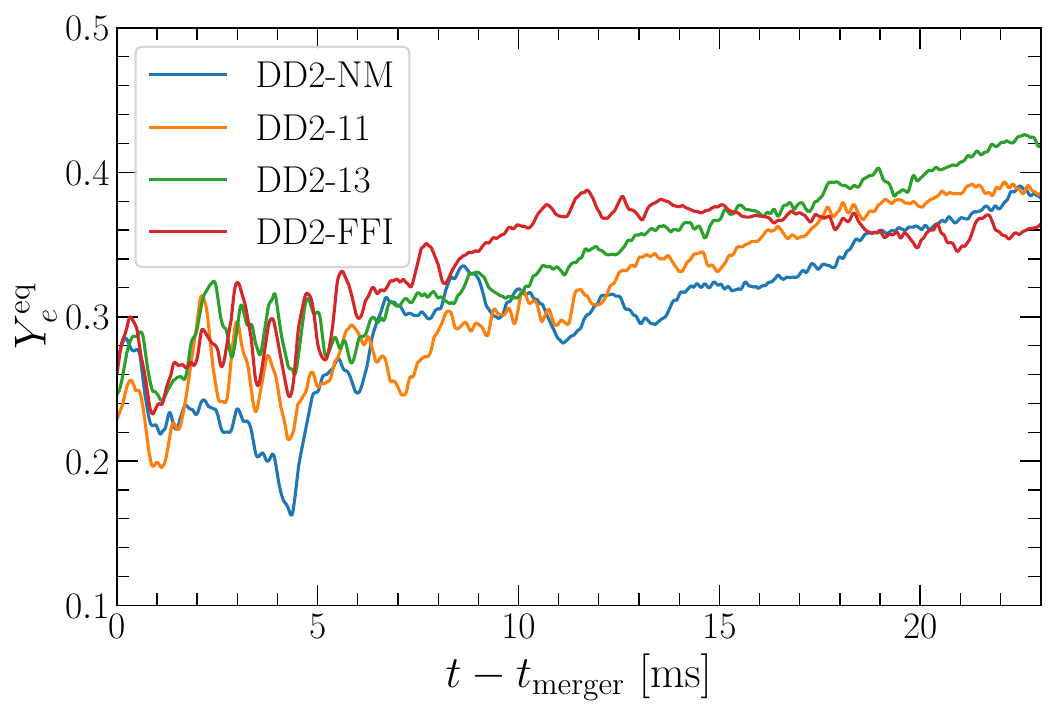}
\includegraphics[width=0.98\columnwidth,keepaspectratio]{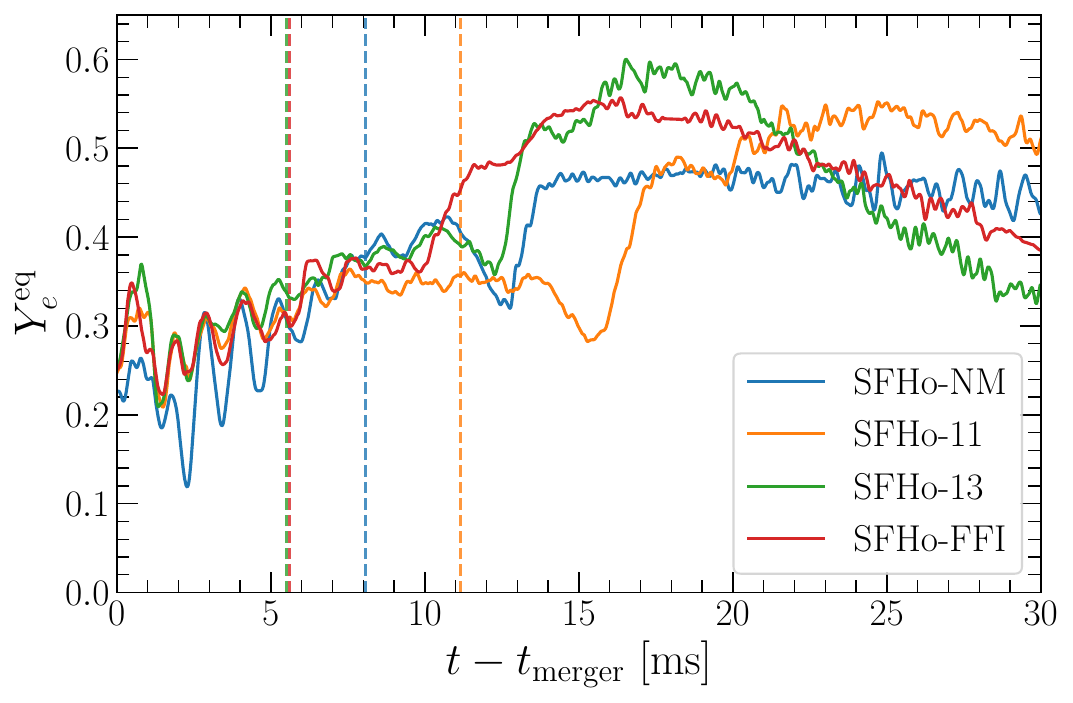}
        \caption{Upper (lower) panels: equilibrium electron fractions for the post-merger ejecta in the four DD2 (SFHo) models. We mark the collapse times for the SFHo runs with dashed lines. The curves are smoothed using convolution with a 0.5 ms square window. The equilibrium $Y_e$ is generally lower in the no-mixing simulation, i.e., the DD2(SFHo)-NM, compared to the mixing models, for most of the time in the DD2 simulations and the time before collapses in the SFHo simulations.}
        \label{fig:eq_ye}
\end{figure}

\begin{figure*}
    \includegraphics[width=2.1\columnwidth]{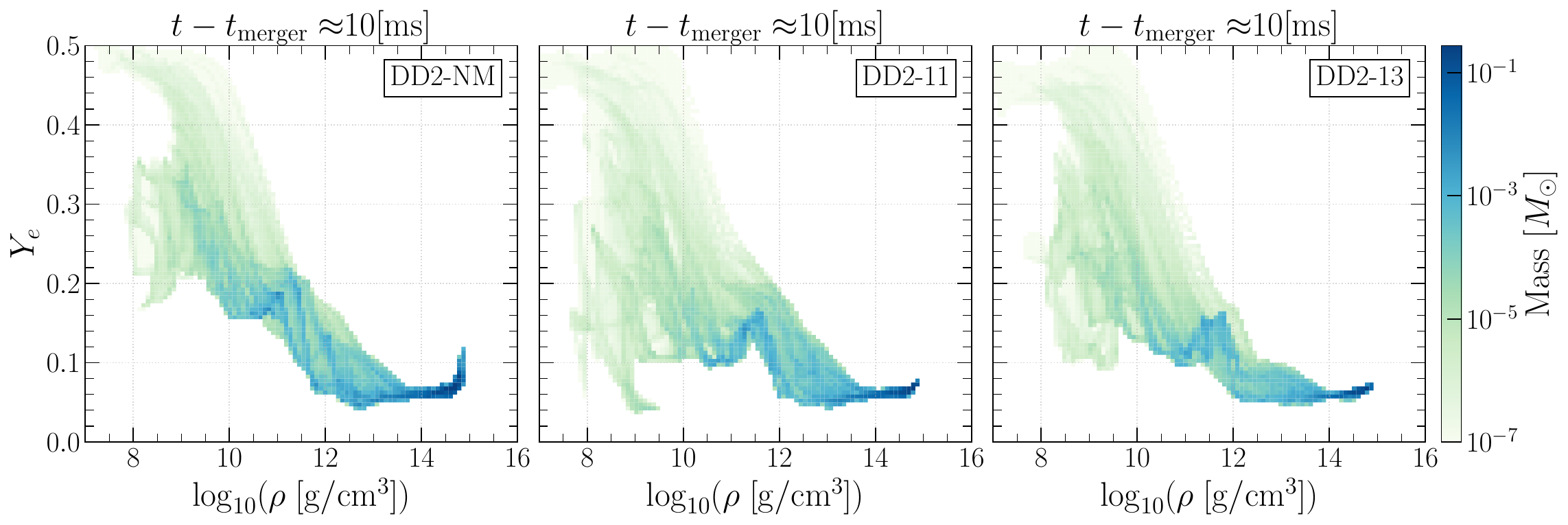}
    \includegraphics[width=2.1\columnwidth]{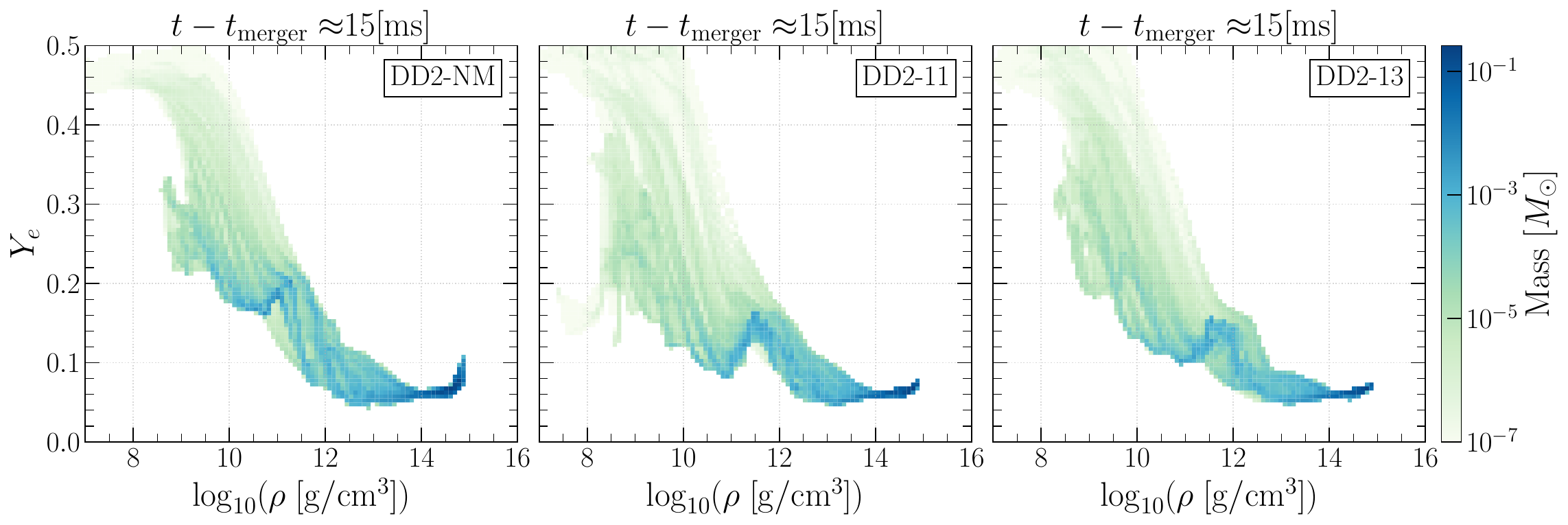}
    \includegraphics[width=2.1\columnwidth]{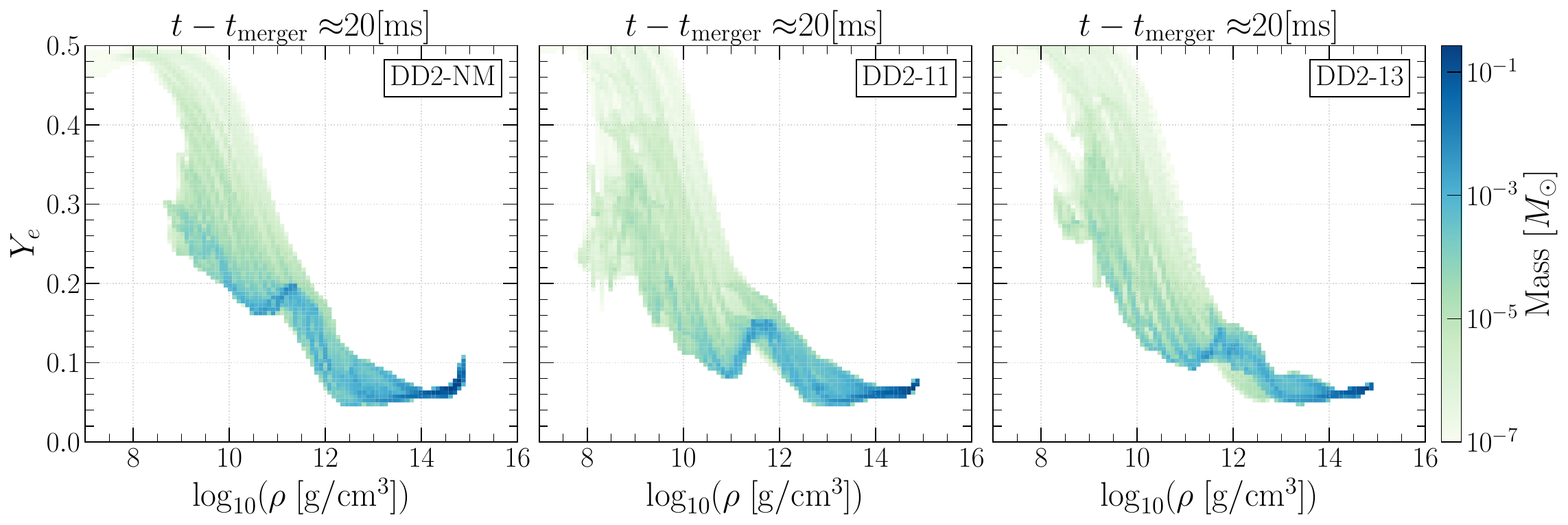}
    \caption{2D histograms of the $Y_e$ versus rest mass density for simulation slices on xz-plane at $\sim$$10,15,20$~ms after merger for DD2-11, DD2-13 and DD2-NM models. Note that we weight the mass in each distribution bin by the radius (x coordinate) of the 2D blocks, so that the results approximate the outcome of the full 3D data. Compared to DD2-NM, the neutrino mixing models DD2-11 and DD2-13 appear to have more neutron rich ejecta (lower $Y_e$ outflow) in the outer disk regions where densities are below $10^{11}$$\mathrm{g/cm}^3$, which approximately corresponds to the density of the electron (anti)neutrino decoupling surfaces. Non-negligible differences can also be observed between the DD2-11 and DD2-13 simulations, which arise in between the $10^{11}$ and $10^{13}$ $\mathrm{g/cm}^3$ regions. These differences indicate the importance of the locations \emph{where} flavor conversions occur, since electron and heavy-lepton type neutrinos decouple at different densities.}
    \label{fig:2dhist}
\end{figure*}

\begin{figure*}
    \includegraphics[width=2.1\columnwidth]{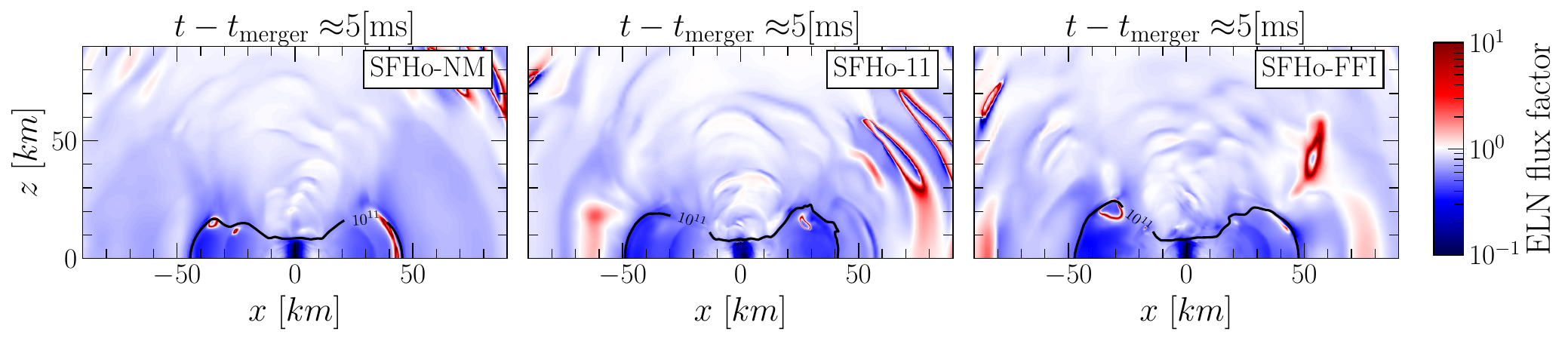}
    \caption{2D ELN flux factors from simulation slices on xz-plane at $\sim$$5$ms after merger for the SFHo-NM, SFHo-11, and SFHo-FFI. The ELN flux factor estimates the ELN crossings, with the ELN flux factors larger than 1 indicating FFIs, based on the ELN flux criterion~\cite{Abbar:2020fcl,Richers:2022dqa}. The overall patterns of the ELN flux factors, especially when viewed across various density ranges, exhibit qualitatively similar behavior among the three models.}
    \label{fig:ELN_flux}
\end{figure*}

\begin{figure*}
\includegraphics[width=2.1\columnwidth]{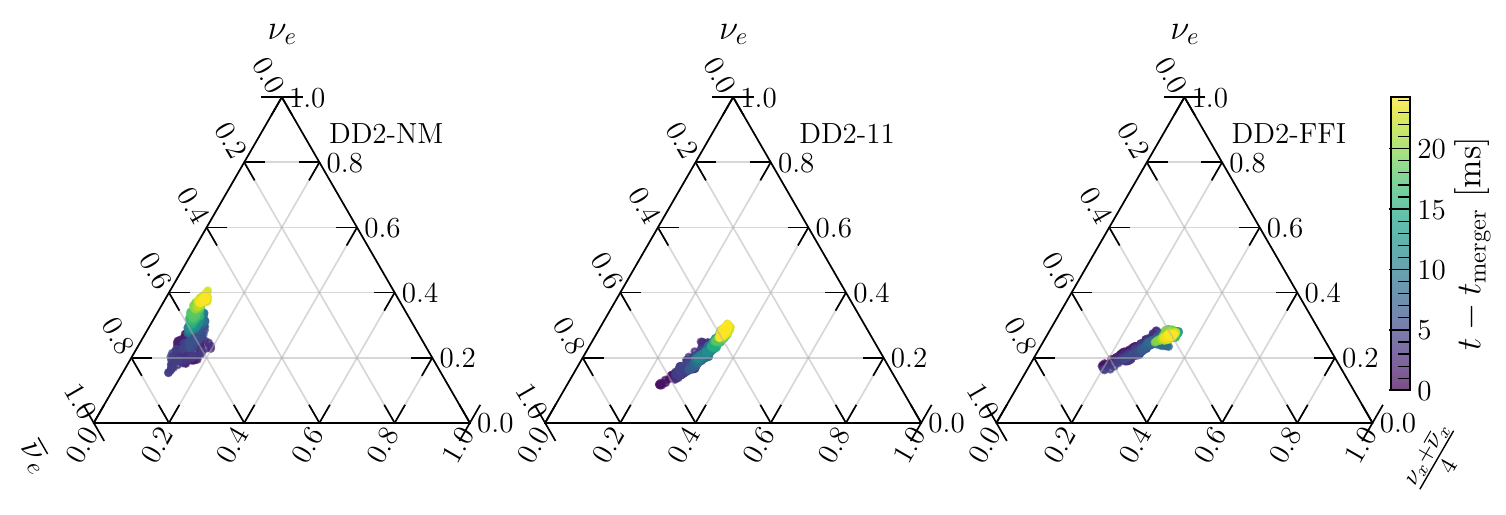}
\includegraphics[width=2.1\columnwidth]{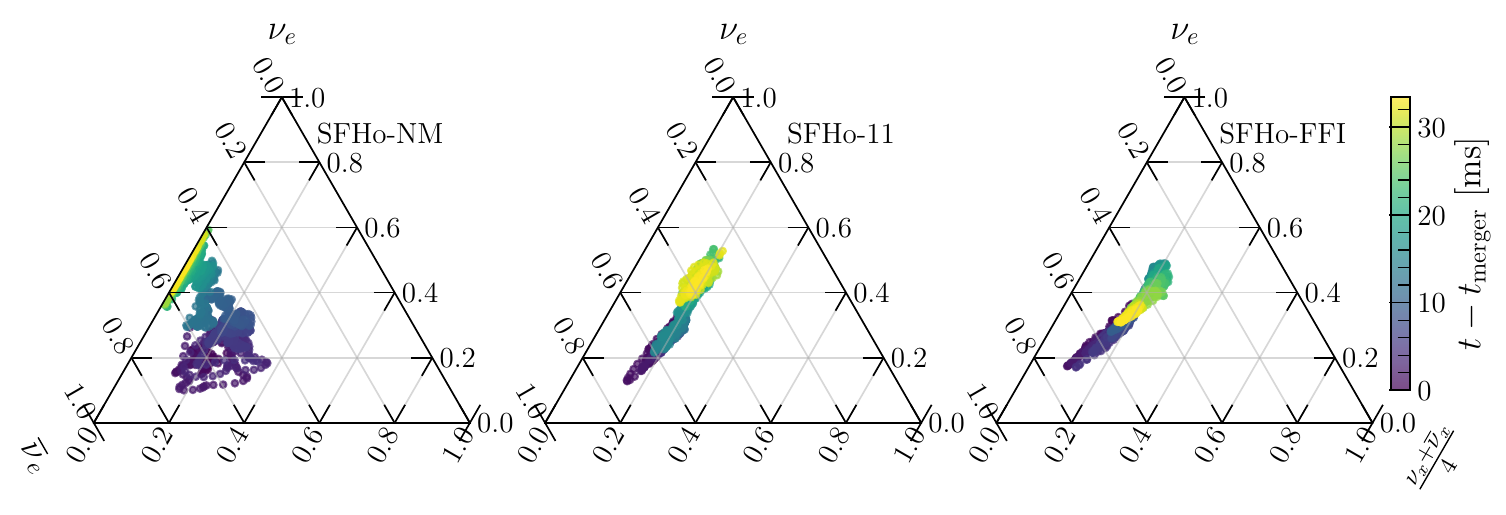}
\caption{Post-merger evolutions of neutrino flavor fractions for the DD2 (SFHo) models in upper (lower) panels. The trajectories represent how the fractions of each neutrino type change over time toward more equilibrated distributions. In models with neutrino flavor conversion, we see consistently higher heavy-lepton neutrino fractions with both DD2 and SFHo EoS's compared to the baseline no-mixing simulations. While the DD2(SFHo)-11 and DD2(SFHo)-FFI models differ mostly in the electron neutrinos' and electron antineutrinos' fractions. FFI motivated models DD2(SFHo)-FFI show less balanced flavor distributions compared to the DD2(SFHo)-11 models, with the final fractions of electron neutrinos and electron antineutrinos being less comparable.}
\label{fig:ternary}
\end{figure*}

\subsection{General dynamics and ejecta properties}
In the DD2 EoS cases, the two stars merge $\sim$$15$~ms from the beginning of the simulations. We keep simulating for an additional $\sim$$25$~ms post-merger to measure the majority of the dynamical mass ejection. With DD2, the remnants are stable against gravitational collapse until the end of the simulations. In the SFHo EoS cases, the stars merge $\sim$25~ms after the beginning of the simulations. The remnant NSs collapse to BHs at $\sim$5-10~ms after merger. The regions inside the BHs are `excised' for both hydrodynamical and neutrino evolutions whenever the horizons are found. The spacetime evolutions are also fixed $\sim$5-10ms after collapses, to avoid numerical instabilities. Because the remnants governed by the SFHo EoS collapse to BHs, their simulations extend to about 35 ms after merger, roughly 10 ms longer than those with the DD2 EoS.

In Fig.~\ref{fig:2d}, we show three snapshots of the simulations at $\sim$5, 10 and 35 ms post-merger for the electron fraction, rest mass densities, electron neutrino and heavy-lepton neutrino number densities on the xy and xz-planes for the SFHo-11 and the SFHo-NM simulations. The $\sim$5 ms post-merger snapshots reflect the phase before the remnant collapses. The $\sim$10 ms post-merger snapshots show the distributions around remnant collapse times. The $\sim$35 ms post-merger snapshots depict the systems at the end of the dynamical phase, when the remnants have entered the accretion phase. The rest mass densities in the two simulations before the remnant collapses look similar. However, due to the differences of the collapse times, as shown in Table~\ref{tab:setup}, the SFHo-11 and SFHo-NM simulations appear to have very different patterns of density distributions after collapse. Such differences are likely not resulting from neutrino flavor conversion effects, but instead arise from slight differences in the time of collapse between the two simulations due to stochasticity. On the other hand, in terms of the composition of the disk and ejecta, we find similar results as those in~\cite{Qiu:2025kgy}. In all stages of the remnant evolutions, the disks in the SFHo-11 mixing model are more neutron rich than those of the SFHo-NM model. Due to flavor conversion effects transforming electron (anti-)neutrinos to heavy-lepton type neutrinos, i.e., $\nu_e, \bar{\nu}_e \rightarrow \nu_x, \bar{\nu}_x$, we also find that the electron (anti)neutrinos are more abundant in the SFHo-NM disks than those of SFHo-11, and vice versa for heavy-lepton neutrinos. 

To better address the impact of flavor conversion on neutrino properties, we extract the neutrino energy fluxes on a $\sim$$295$km radius sphere away from the center of the simulation domain, and hence obtain the neutrino luminosities for the total and individual species. In Fig.~\ref{fig:lumi} we plot the luminosity evolutions for both DD2 and SFHo runs. The total neutrino luminosities are typically larger in the mixing simulations, compared to the no-mixing cases, for most of the evolutions. The main exception occurs after BH formation in the SFHo cases, where the outcomes depend on stochasticity in the collapse time and the mass of the remnant disk. In the case of a massive remnant (i.e. in DD2 simulations and until BH collapse in the SFHo simulations), the total neutrino luminosity is the largest in the DD2(SFHo)-13, compared to the other mixing models. This is due to the fact that the region where flavor conversion occurs is larger in the DD2(SFHo)-13, compared to the DD2(SFHo)-11 and the DD2(SFHo)-FFI models, since it includes the inner disk, with densities between $10^{11}$$\mathrm{g/cm}^3$ and $10^{13}$$\mathrm{g/cm}^3$, and FFI-stable regions. Inside these disk layers, where electron (anti)neutrinos are still significantly emitted by neutrino-matter processes, heavy-lepton neutrinos are produced through flavor conversions and then propagate more freely outwards, and possibly convert back to electron (anti)neutrinos only in the outer disks~\cite{Qiu:2025kgy}. This results in a more efficient remnant cooling. Conversely, the total luminosity observed in the DD2-11 simulation is closer to the DD2-NM case, since oscillations occurring below $10^{11} {\rm g/cm^3}$ do not significantly affect the overall remnant cooling.

For individual species, due to neutrino flavor conversions $\nu_e, \bar{\nu}_e \rightarrow \nu_x, \bar{\nu}_x$, we see in both EoS cases higher luminosities of electron (anti-)neutrinos in the DD2(SFHo)-NM simulation, compared to the DD2(SFHo)-11, the DD2(SFHo)-FFI and the DD2(SFHo)-13 simulations, and vice versa for heavy-lepton neutrinos. In particular, we see DD2-13 simulation has up to $400\%$ larger heavy-lepton neutrino luminosity than the DD2-NM simulation, up to $100\%$ more than the DD2-11 simulation, and up to $30\%$ more than the DD2-FFI simulation. However, for the SFHo runs, we do not observe a clear hierarchy, due to differences in the collapse times, as shown in Table~\ref{tab:setup}. At later times, the SFHo-11 has higher luminosities in all species, while the other mixing models are generally comparable. Despite the hierarchy being not very clear in the SFHo runs, we emphasize that, in general, mixing significantly changes the absolute values of the luminosities and the hierarchy with regard to the no-mixing case, such that the luminosities become qualitatively much closer among different flavors. To sum up, the results suggest that flavor conversions of $\nu_e, \bar{\nu}_e \rightarrow \nu_x, \bar{\nu}_x$ have significant impacts on the resulting neutrino luminosities. In particular, neutrino mixing changes the luminosity hierarchy with respect to the no-mixing case. Moreover, the ample range of luminosities observed for the same binary configuration and EoS suggests that both \emph{where} and \emph{how} neutrino flavor conversions occur are crucial in determining the neutrino luminosities.

Additionally, in Fig.~\ref{fig:mean_energy} we show the neutrino mean energy measured on a $\sim$$295$km radius sphere for $\nu_e$, $\bar{\nu}_e$, and $\nu_x$ . Comparing to the no-mixing case, neutrino mixing models have higher electron (anti)neutrino mean energies and lower heavy-lepton neutrino mean energy, in both DD2 and SFHo runs. The mean energies of different flavors are very similar in the mixing models, except that $\nu_e$ seems a bit smaller, which could be due to the fact that high energy $\nu_e$ are absorbed more easily by the neutron rich material and the spectrum tends to become a bit softer. When flavor conversion effects are absent, heavy-lepton neutrinos generally have higher mean energies compared to the electron (anti)neutrinos since they decouple in the deeper regions of the disk. If flavor transformations occur in the inner disk, i.e., in the DD2(SFHo)-13 runs, $\nu_e$ and $\bar{\nu}_e$ can escape more easily, which then increase the net luminosity, as we see in Fig.~\ref{fig:lumi}. Effectively, the changes in neutrino mean energy spectra means that neutrino flavor conversion pushes the electron flavor decoupling region inward and the heavy-lepton flavor decoupling region outward. For the other mixing models, because the flavor conversions are activated outside the neutrino decoupling surfaces, the shifts of mean energies only represent an exchange of energy of different flavors of neutrinos through flavor transformations. 

We report in Table~\ref{tab:setup} some of the main ejecta properties of all the simulations. Note that we keep track of the mass ejection and compute the unbound mass based on geodesic criterion, i.e., the matter has to have the time component of its 4-velocity smaller than $-1$. We see that the runs with BH remnants have larger total ejecta mass, and faster ejecta, due to the softer EoS. We also show the ejecta electron fraction profiles and nucleosynthesis yields for both DD2 and SFHo EoSs with 4 models in Fig.~\ref{fig:ejecta}. In particular, neutrino flavor conversions make the ejecta more neutron rich in the DD2(SFHo)-11, DD2(SFHo)-13, and DD2(SFHo)-FFI simulations compared to the DD2(SFHo)-NM case. Remarkably, the very neutron-rich ejecta fraction ($Y_e < 0.15$, roughly the threshold for actinides production) can differ by a factor of 3 between the DD2-11 and DD2-NM models, while all the neutron-rich ejecta fraction ($Y_e < 0.25$, roughly the threshold for lanthanides production) also differs by about a factor of 3. These values likely represent an upper bound on the discrepancy between neutrino-mixing and no-mixing scenarios. In contrast, the same comparison with the SFHo EoS (SFHo-11 vs. SFHo-NM) shows only about a $\sim100\%$ difference for the $Y_e < 0.15$ ejecta, and roughly a $\sim50\%$ difference for the $Y_e < 0.25$ ejecta. The relative order between DD2(SFHo)-11 and DD2(SFHo)-13 is reversed for the two different EoSs runs. And the amplification of $Y_e < 0.15$ ejecta in the FFI models is smaller than that of simple density-threshold neutrino mixing models. This is because in the DD2(SFHo)-FFI models, their instability criterion does not flag the entire volume to have neutrino mixing. For the nucleosynthesis yields, the DD2-11 simulation produces about an order of magnitude more lanthanides and heavier elements than the DD2-NM simulation, when the yields are normalized to $A=85$. In the SFHo simulations, the difference is smaller but still reaches up to a factor of five between the SFHo-11 and SFHo-NM relative heavy element yields. \reply{In particular, we find quantitative agreement with~\cite{Just:2022flt} regarding the heavy-element yields in the SFHo-FFI model, with the difference between the mixing and no-mixing cases being around 100\%, despite our ejecta properties being measured at earlier, closer-to-merger timescales.} Overall, the amount of low-Ye ejecta correlates with the amount of heavy element production.

We also show the equilibrium $Y_e$~\cite{Qian:1996xt} of the ejecta in Fig.~\ref{fig:eq_ye}, by assuming
\begin{equation}
Y_{e}^\mathrm{eq} \approx\left(1+\frac{L_{\bar{\nu}_e} \langle\epsilon_{\bar{\nu}_e}\rangle}{L_{\nu_e} \langle\epsilon_{\nu_e}\rangle}\right)^{-1}.
\end{equation}
The fact that the ejecta are more neutron rich in the mixing models, as visible in Fig.~\ref{fig:ejecta}, but their equilibrium $Y_e$ is higher suggests that the effect of flavor conversions is mostly to make the time scale of equilibration longer by reducing the $\nu_e$ luminosity by a factor of $2\sim10$ (see in Fig.~\ref{fig:lumi}). Since the final electron fractions of the ejecta in the mixing models are determined by the competition between the equilibration and the expansion time scales, as well as by the combined action of both thermodynamical and flavor conversion effects.

To characterize the properties of the remnant and of the disk in the different DD2 models and their evolution, we compute mass-weighted histograms in the rest mass density-$Y_e$ plane. We consider xz slices extracted at different times and assume axial symmetry (which is a good approximation once the disk has formed).
The results are shown in Fig.~\ref{fig:2dhist}. In the DD2-11 and DD2-13 simulations, the histograms show larger amounts of mass in the low-$Y_e$, low-density regions compared to the DD2-NM simulation, consistent with the increase of low-$Y_e$ ejecta in Table~\ref{tab:setup}. This reflects the enhanced low-$Y_e$ outflow caused by reduced charged current weak interactions following flavor conversions from electron neutrinos to heavy-lepton neutrinos. We also observe non-negligible differences between the DD2-11 and DD2-13 simulations. We see that the DD2-13 has slightly lower $Y_e$ in regions with densities between $10^{11}$ and $10^{13} \mathrm{g/cm^3}$. This indicates the effects of the locations where flavor conversions occur. Over time, the low-density outflows gradually shift toward higher $Y_e$, though more so in the no-mixing models than in the mixing models.

\subsection{Neutrino flavor evolution}
We show snapshots on xz-planes of the ELN flux factors (defined in Eq.\ref{eq:ELN_flux}) at $\sim$5 ms post-merger for the SFHo-NM, SFHo-11 and SFHo-FFI simulations in Fig.~\ref{fig:ELN_flux}. We find that in all the models, there are scattered regions that would be unstable to FFIs. The magnitudes of the ELN flux factors are slightly smaller in the two mixing models, i.e., the SFHo-11 and SFHo-FFI than those in the SFHo-NM. The precise locations of the instabilities differ, likely due to stochasticity across different simulations. However, the overall patterns of the ELN flux factors exhibit qualitatively similar behavior among the three models. The instabilities in polar regions could possibly be explained by the findings in~\cite{Nagakura:2025hss}, which discussed correlations between the neutrino chemical potential and the ELN crossing, and suggested that the instabilities could be generated by $\nu_e$ contamination. Given that the FFI models explicitly target regions where the ELN flux factor is above 1, one would expect the neutrino flavor mixing to remove such instabilities by pushing them towards flavor equilibrium~\cite{Zaizen:2022cik}, as the prescriptions for equilibrium states we consider in the BGK model decrease the ELN flux factor. However, our results show that flavor conversions do not necessarily eliminate the instabilities that trigger them. The fact that instabilities remain indicates that the dynamics generating the instabilities must pump the distribution up to unstable levels very rapidly, prevent the system from settling into a stable flavor equilibrium.

In Fig.~\ref{fig:ternary}, we show ternary diagrams illustrating the post-merger evolution of neutrino luminosity fractions measured on a sphere at ${\sim}$295~km from the center of the simulation domain. In all models, the flavor composition during merger initially shows a dominance of electron antineutrinos. Over time, the compositions move toward equilibrium points on the diagram and eventually stabilize. In models with neutrino flavor conversion effects (DD2(SFHo)-11 and DD2(SFHo)-FFI), the heavy-lepton neutrino fractions are higher than those in the DD2(SFHo)-NM models, due to flavor conversions from electron neutrinos. Differences between the mixing models mainly appear in the electron neutrino and electron antineutrinos fractions, in the SFHo models. The FFI models have roughly the same fraction of heavy-lepton neutrinos, but more electron antineutrinos and less electron neutrinos, as compared to the DD2(SFHo)-11 models. Overall, the neutrino flavor patterns seems to be quite robust, at least within the models we have explored. Nevertheless, these results highlight the dynamics of neutrino transport and the potential impact of different flavor equilibrium prescriptions.

\begin{figure}
\includegraphics[width=0.98\columnwidth]{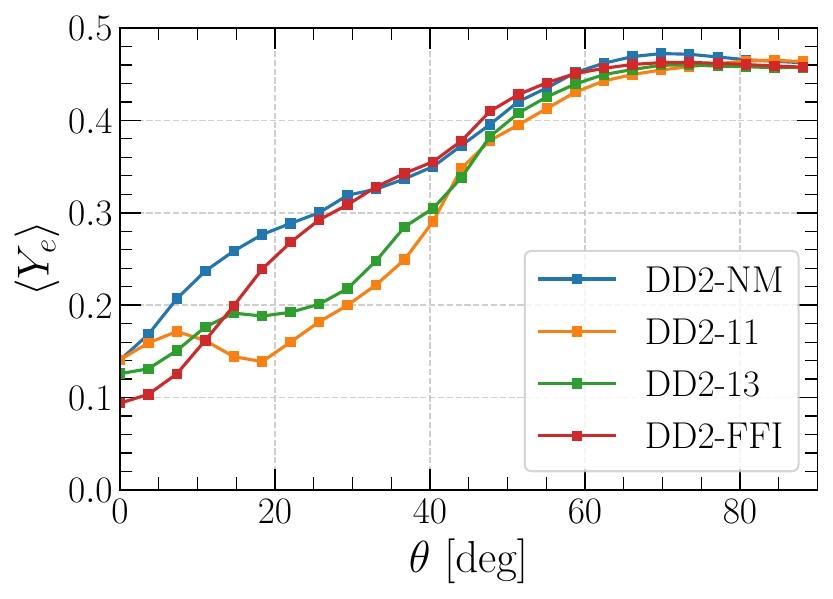}
\includegraphics[width=0.98\columnwidth]{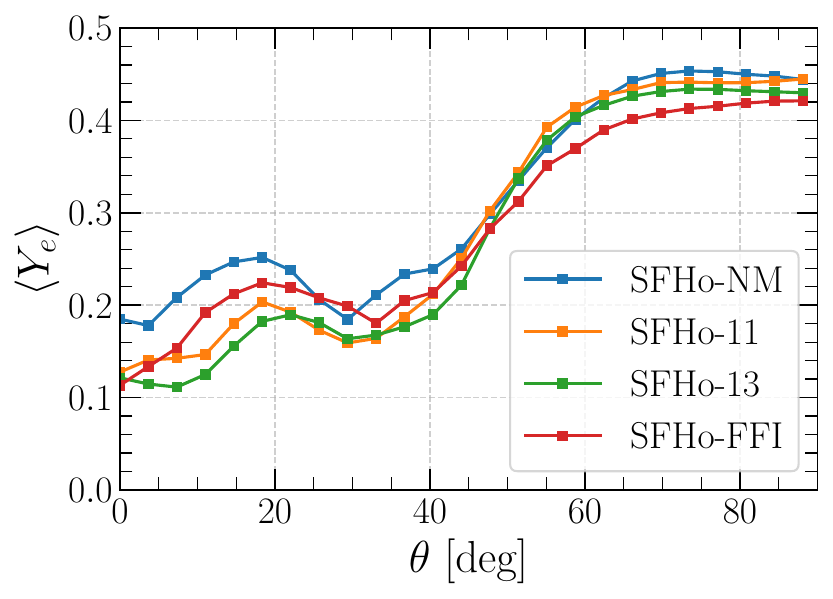}
    \caption{Upper (lower) plot shows the ejecta average electron fraction vs. $\theta$ in the DD2 (SFHo) models. Note that here we define $\theta$ simply as the latitude. We see that the differences between mixing and no-mixing models are mainly in the equatorial plane and intermediate latitudes ($\theta<45^\circ$). The average $Y_e$ increases almost monotonically with the latitude for the DD2 runs. While for the SFHo runs, the average $Y_e$ of ejecta shows a small bump around $\theta\approx20^\circ$, which is likely sourced from shock heating.} 
    \label{fig:theta}
\end{figure}

\subsection{$Y_e-\theta$ profiles}
In Fig.~\ref{fig:theta} we show the average electron fractions for the ejecta as a function of the latitude $\theta$, for both the DD2 and SFHo runs. For the DD2 models, the ejecta average $Y_e$ increases almost monotonically with latitude. While for the SFHo runs, the average $Y_e$ of ejecta shows a small bump around $\theta\approx20^\circ$, which is likely associated with shock heating. The main differences between the no-mixing and mixing runs with the DD2 EoS occur near the equatorial plane and at intermediate latitudes ($\theta<45^\circ$), with their average $Y_e$ differing by up to $0.1$. In the SFHo runs, the contrast between mixing and no-mixing models is likewise most pronounced in the near-equatorial ejecta, with little variation in the polar regions. The only exception is the SFHo-FFI model, which exhibits up to $\sim0.05$ lower average $Y_e$ than the SFHo-NM model at higher latitudes ($\theta>45^\circ$). The differences in the $Y_e-\theta$ profiles between mixing and no-mixing models suggest that neutrino flavor conversions have larger impacts on ejecta near the equatorial plane and at intermediate latitudes, than ejecta in the polar regions. \reply{This is because the equatorial ejecta expand on timescales shorter than the weak equilibration timescale, whereas the polar ejecta expand more slowly and experience stronger neutrino irradiation, especially in the longer-lived remnant case. Therefore, the changes in neutrino luminosities and mean energies caused by flavor conversion have a more limited effect on the electron-fraction evolution of the polar ejecta.}

\begin{figure}
\includegraphics[width=0.98\columnwidth]{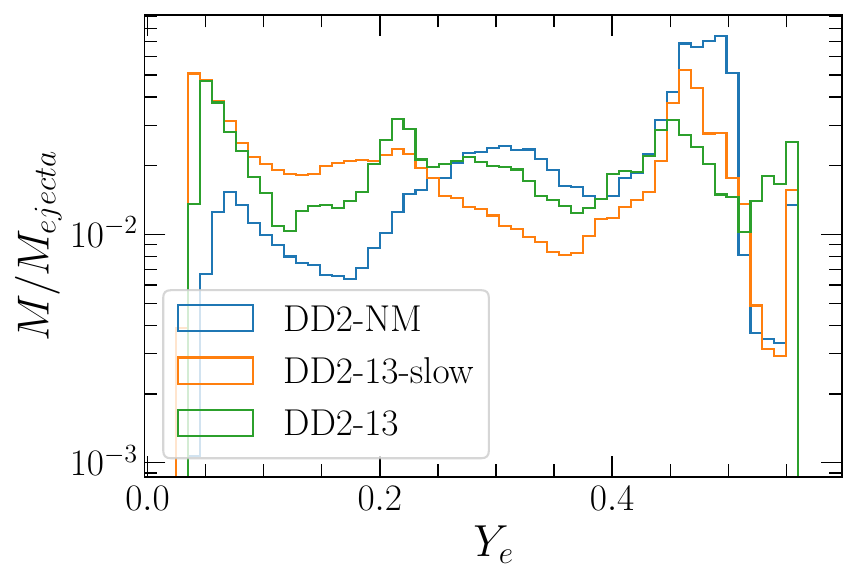}
    \caption{Histograms of the electron fraction distributions of the ejecta for the DD2-13-slow and DD2-13 simulations. The former uses ${\sim}50$~ns relaxation time for flavor conversions, while the latter uses ${\sim}0.5$ ns (which is what we use for all the other mixing models throughout this work). The DD2-13-slow ejecta is slightly more neutron rich than that of the DD2-13. While both of them are much more neutron rich than the ejecta of the DD2-NM.} 
    \label{fig:tau}
\end{figure}

\begin{figure}
\includegraphics[width=0.98\columnwidth]{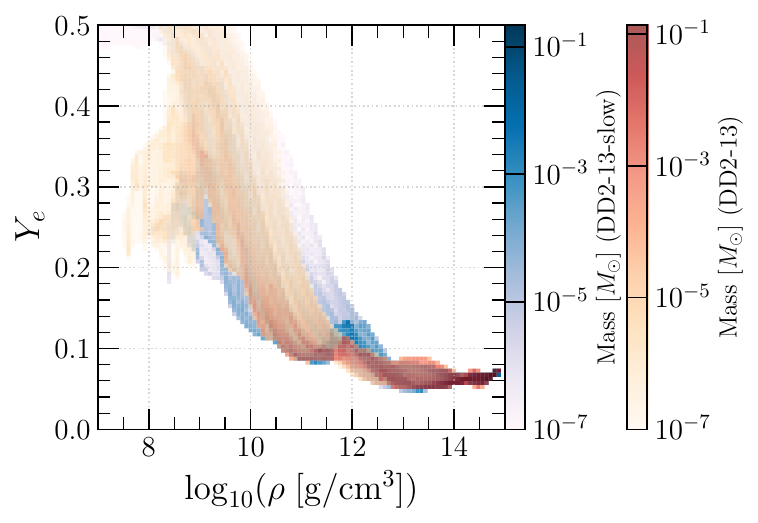}
    \caption{2D histograms of $Y_e$ versus rest mass density for simulation slices on xz-plane at $\sim$$20$ ms after merger for the DD2-13-slow and DD2-13 runs. Note that we weight the mass in each distribution bin by the radius (x coordinate) of the 2D blocks, so that the results approximate the outcome of the full 3D data. Major differences appear in the inner disk regions at densities between $10^{11}$ and $10^{13}$ $\mathrm{g/cm}^3$, where we see higher electron fractions in the DD2-13-slow simulation than in the DD2-13 simulation. We can also observe differences in low density regions at densities below $10^{11}$$\mathrm{g/cm}^3$, where the $Y_e$ is generally lower in the DD2-13-slow simulation than that in the DD2-13 simulation. 
    } 
    \label{fig:2dhist_tau}
\end{figure}

\begin{figure}
\includegraphics[width=0.98\columnwidth]{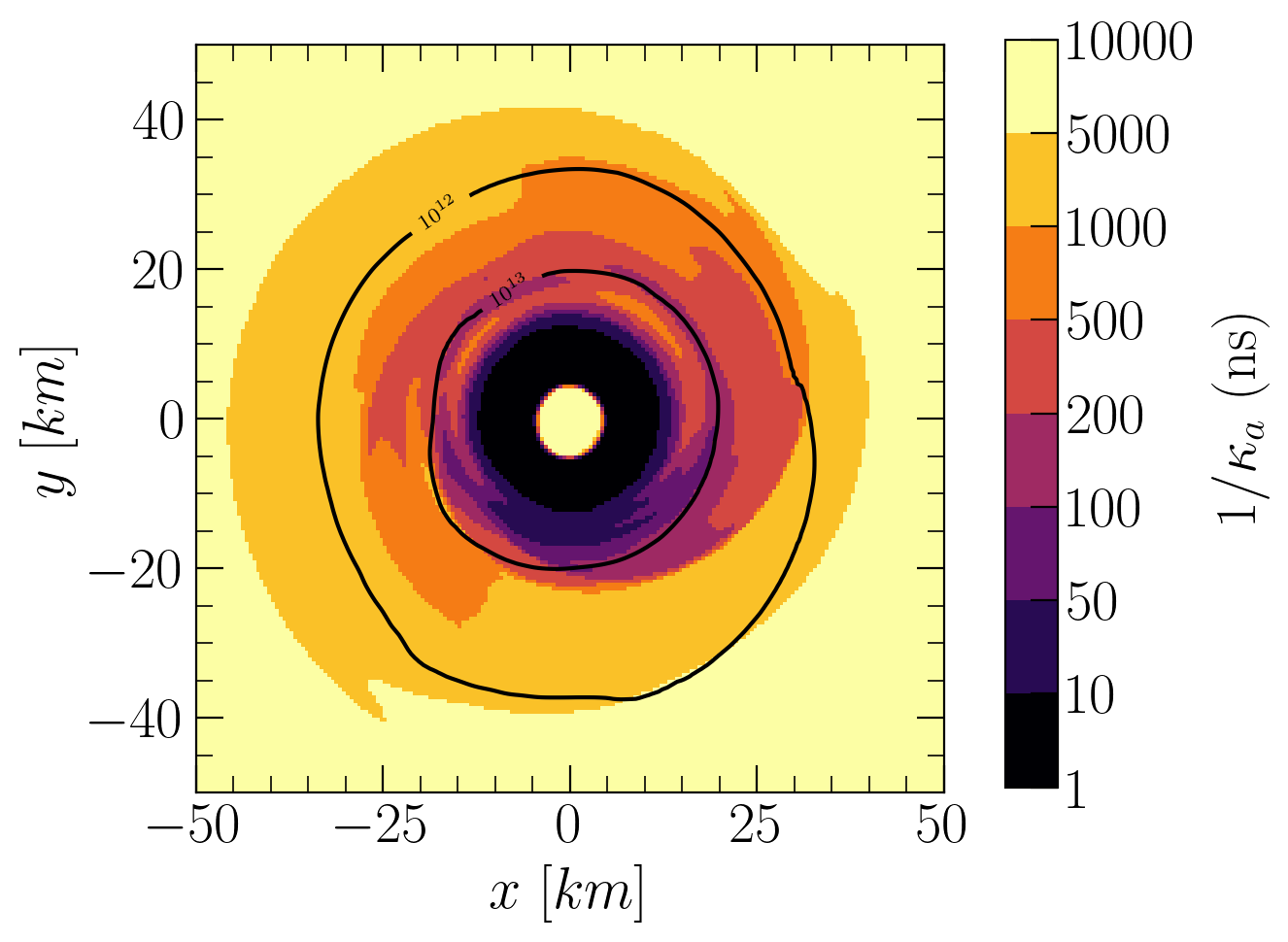}
    \caption{Electron-neutrino absorption times on the $xy$-plane at $\sim$$20$ ms after merger for the DD2-13 run. Around the $10^{13},\mathrm{g/cm}^3$ density contour, these absorption times—which approximate the thermal equilibration times of electron neutrinos—typically fall in the range of 100–200 ns. Notably, this timescale is comparable to the $50\,\mathrm{ms}$ relaxation time adopted in the DD2-13-slow model and much slower than the $0.5\,\mathrm{ns}$ relaxation time adopted in the other simulations.} 
    \label{fig:opacity}
\end{figure}

\subsection{Effect of neutrino relaxation time}
We also vary the relaxation times for the density dependent mixing model DD2-13. We consider a new model with a relaxation time of $50$ ns, in which flavor conversions happen on a time scale similar to our evolution time step of $40$ ns, denoted as DD2-13-slow. In contrast, our default DD2-13 model assumes a $0.5$ ns relaxation time. Also note that we have only considered the slow relaxation time scenario at LR resolution. In Fig.~\ref{fig:tau}, we see that the ejecta of the DD2-13-slow binary seems to be slightly more neutron rich than those in the DD2-13 binary, with relative difference being $40\%$ in the $M^{\mathrm{ej}}_{Y_e<0.15}/M^{\mathrm{ej}}_\mathrm{total}$ and $20\%$ in the $M^{\mathrm{ej}}_{Y_e<0.25}/M^{\mathrm{ej}}_\mathrm{total}$ as shown in Table~\ref{tab:setup}. Such differences are likely sourced from the shift of the equilibrium $Y_e$ towards higher values for DD2-13-slow in the inner disk regions, which appears in Fig.~\ref{fig:2dhist_tau}. This happens because the $50$ ns relaxation time for flavor conversions is comparable to thermal equilibration time for density around $10^{13}$$\mathrm{g/cm}^3$, as we see in Fig.~\ref{fig:opacity}, resulting in a competition between the two processes. In the end, the material at lower densities ends up with lower electron fraction in the outer disk regions for DD2-13-slow comparing to those in DD2-13, as we can see in Fig.~\ref{fig:2dhist_tau}. This also highlights that the details of the subgrid model, like the relaxation times of the flavor conversions, have nontrivial effects on the ejecta properties. A proper treatment of flavor transformation could have similarly large quantitative discrepancies from any ad-hoc subgrid model.


\begin{figure}
\includegraphics[width=0.98\columnwidth]{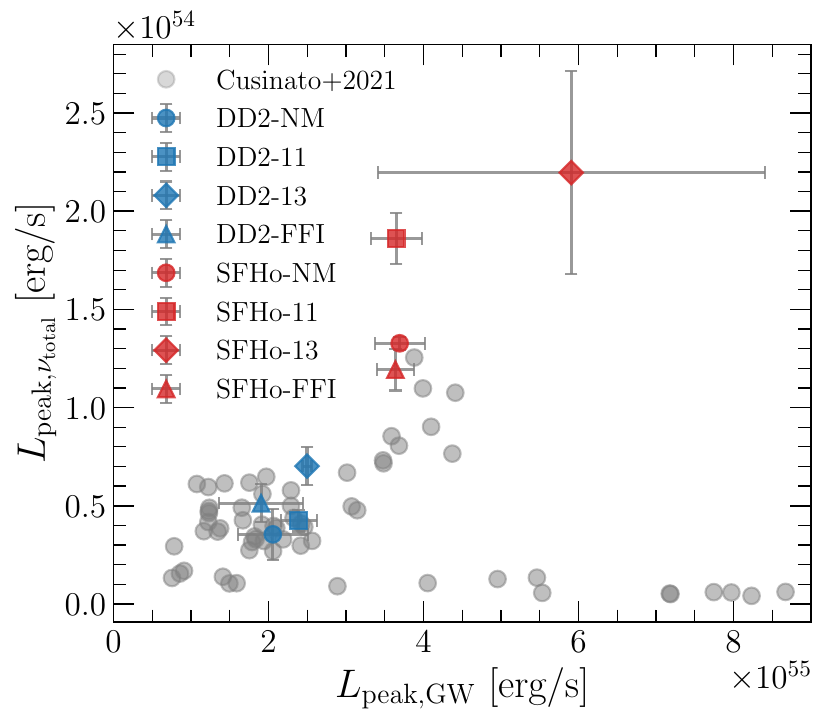}
    \caption{Peak total neutrino luminosities $L_{\text {peak}, \nu}$ vs. peak GW luminosity $L_{\text {peak, GW }}$ for our 8 models. Also shown is data for the M0 simulations considered in~\cite{Cusinato:2021zin}, for reference. The error bars are estimated from differences between the SR and LR simulations. The peak luminosities in both GWs and neutrinos are higher for the BH forming runs. When compared the results among the same remnant type, we see that the DD2(SFHo)-13 are the brightest for both messengers. The rest of the models have comparable peak GW luminosities, when taking into consideration the error bars. For the peak total neutrino luminosities, the SFHo-11 model ranks second, followed by SFHo-NM and SFHo-FFI, which are consistent within their uncertainties. Similarly, the DD2 models exhibit comparable peak luminosities, except for DD2-13. 
    } 
    \label{fig:peak_lumi}
\end{figure}

\subsection{Neutrino and GW luminosities}
In ~\cite{Qiu:2025kgy}, we found the evolution of GW energies to differ for different models, where neutrino mixing models generally have higher GW energies. Here, we show in Fig.~\ref{fig:peak_lumi} the GW and neutrino peak luminosities for all our models. The grey scatter points in the background show reference data from the M0 simulation results discussed in~\cite{Cusinato:2021zin}. Excluding the grey points near the x-axis—corresponding to mergers that promptly collapse to BHs—there is a clear correlation between the GW and neutrino peak luminosities, highlighting the link between neutrino transport and merger dynamics. Within our models, the DD2(SFHo)-13 runs exhibit the highest neutrino peak luminosities. This arises because flavor conversions into heavy-lepton neutrinos in the inner disk enhance neutrino escape, as discussed earlier. The increased flavor mixing in these runs also leads to more compact remnants, resulting in stronger GW emission. The other three models have comparable GW peak luminosities when taking into account their error bars. For the neutrino peak luminosities, the SFHo-11 is the second highest, followed by the SFHo-NM and then the SFHo-FFI. The latter two are comparable considering their uncertainties. Comparable peak neutrino luminosities are also observed among the DD2 models, except for DD2-13. This suggests that flavor transformations can not strongly affect the neutrino peak luminosities, unless they are able to occur at very high densities. 


\section{Conclusion}
We studied the role of neutrino flavor conversions in BNS mergers resulting in either a BH or NS as remnant. We used a BGK operator to model parameterized neutrino flavor conversion possibly produced by quantum many-body effects, beyond standard-model physics, or fast flavor instabilities, with varying assumptions for relaxation times and flavor equilibrium states. Across all simulations, we comprehensively analyzed the merger dynamics, ejecta compositions, neutrino flavor evolutions, nucleosynthesis yields, and the associated multi-messenger signatures. 

In agreement with earlier work~\cite{Qiu:2025kgy,Li:2021vqj,Just:2022flt}, we find that mixing scenarios drive the matter toward more neutron-rich conditions compared to no-mixing models. The neutrino flavor transformations of electron (anti)neutrinos into heavy-lepton types in the inner disk regions boost the total neutrino luminosities at larger radii. While species-dependent hierarchies vary across the EoS and collapse outcomes, we generally find heavy-lepton luminosities amplified in the mixing models relative to the no-mixing case, and vice versa for the electron (anti)neutrinos. Flavor conversions raise the mean energies of electron (anti)neutrinos and lower those of heavy-lepton neutrinos, effectively shifting the decoupling surface of electron-flavor neutrinos inward and that of heavy-lepton neutrinos outward. These effects carry over into the ejecta, where flavor conversions increase the low-$Y_e$ material by up to $300\%$ in the runs that have longer-lived NS remnants, translating into significantly larger relative r-process yields (up to a factor of 10). Note that the ejecta differences observed significantly exceed the reported $10\%$ to $30\%$ numerical uncertainties estimated from comparisons between M1 and Monte Carlo neutrino transport schemes~\cite{Foucart:2024npn}. \reply{While the enhancement in heavy-element production in the FFI model is on the order of 100\%, which is consistent with the magnitude reported in~\cite{Just:2022flt}.} These differences are mainly driven by neutrino interactions in low-density regions, where neutrinos are less constrained by the surrounding matter and can oscillate more freely in flavor space.

All our models develop sparse regions unstable to FFIs and the patterns look qualitatively similar to those in~\cite{Nagakura:2025hss}. Notably, we find that the FFI-targeted flavor equilibration does not simply erase the instabilities. Unlike the more detailed, but local, calculations present in the literature, our global simulations allow us to capture non-local transport effects and dynamical changes in the neutrino distributions, which may contribute to the regeneration of instabilities. We emphasize that this conclusion should be validated with more sophisticated treatments of FFIs, particularly in accurately identifying their onset conditions and equilibrium states. While all simulations evolve toward flavor equilibrium points, the mixing models in general shift the flavor balance by enhancing heavy-lepton neutrino fractions at the expense of electron-type species. The density-11 mixing models, in particular, tend to produce more balanced fractions of $\nu_e$ and $\bar{\nu}_e$ compared to the FFI models, for short-lived remnant. 

The angular dependence of the ejecta composition indicates that neutrino flavor conversions predominantly affect the low- and mid-latitude outflows, while their impact remains minimal in the polar regions. The near-equatorial and intermediate-latitude ejecta show the strongest sensitivity to mixing, resulting in a reduction of the average $Y_e$ by up to ${\sim}0.1$. In contrast, the polar ejecta generally maintain consistently higher values regardless of flavor mixing. These trends suggest that the enhanced production of heavy elements in the mixing models is primarily driven by equatorial ejecta, in contrast to the behavior reported in~\cite{Lund:2025jjo}. This geometric dependence highlights that the nucleosynthetic signatures of flavor conversions are inherently multidimensional and strongly shaped by the spatial distribution of mass ejection.

The quantitative ejecta properties also depend sensitively on the timescale of flavor equilibration. Comparing the neutrino mixing simulations with two different relaxation times, we find up to 40\% relative differences in the fraction of very neutron-rich ejecta ($Y_e < 0.15$) and $\sim$20\% relative differences in the total low-$Y_e$ mass fraction. These variations occur despite the equilibrium $Y_e$ in the inner disk having higher values. This result shows that flavor conversions interact with matter equilibration, and the competitions of the two across different density ranges quantitatively shape the ejecta composition and r-process yields.

Neutrino flavor conversions also modify the neutrino and GW peak luminosities, especially if allowed to occur in the high-density inner regions of the accretion disk. Flavor transformations enhance the neutrino escape rate from the disk, leading to a more compact remnant. We identify a positive correlation between neutrino and GW peak luminosities, highlighting the close coupling between neutrino transport physics and the bulk dynamics of the merger.

As one of the first explorations of flavor conversion effects in neutron star mergers, our study emphasizes both the potential significance and the current uncertainties of such processes. We call for future efforts to build a more complete picture. A more comprehensive treatment of neutrino transport—including the effects of magnetic fields~\cite{Tambe:2024usx,Most:2025kqf}, muonic interactions~\cite{Pajkos:2024iml,Ng:2024zve}, inelastic neutrino-lepton scatterings~\cite{Cheong:2024cnb,Chiesa:2024lnu}, and pair processes—will be essential, as these missing ingredients can all have quantitative impacts. Equally important are theoretical advances in understanding the precise locations, timescales, and nonlinear outcomes of flavor instabilities, e.g.~\cite{Richers:2024zit,Liu:2025tnf,Nagakura:2025hss,Laraib:2025uza,Liu:2025muc} (see also a review in~\cite{Johns:2025mlm}), which are critical for accurately modeling flavor conversions in global merger simulations.

\begin{acknowledgements}
YQ, DR, and MB were supported by the U.S. Department of Energy, Office of Science,
Division of Nuclear Physics under Award Number(s) DE-SC0021177 and DE-SC0024388.
DR acknowledges support from the Sloan Foundation, from the National Science
Foundation under Grants No. PHY-2020275, AST-2108467, PHY-2116686, and
PHY-2407681. SR acknowledges support from the National Science Foundation under Grant No. PHY-2412683. 
FMG and AP are supported by the European Union under NextGenerationEU, PRIN 2022
Project No.~2022KX2Z3B.
FMG and AP also acknowledge the EuroHPC Joint
Undertaking for awarding this project access to the EuroHPC supercomputer LUMI, hosted by CSC (Finland)
and the LUMI consortium through a EuroHPC Extreme
Scale Access call (EHPC-EXT-2022E01-046).
Simulations were performed on TACC's Frontera (NSF LRAC allocation
PHY23001), on NERSC's Perlmutter, on CSC-LUMI (EHPC-EXT-2022E01-046), and on the Pennsylvania State University’s
Institute for Computational and Data Sciences’ Roar supercomputer. This research
used resources of the National Energy Research Scientific Computing Center, a
DOE Office of Science User Facility supported by the Office of Science of the
U.S.~Department of Energy under Contract No.~DE-AC02-05CH11231.
\end{acknowledgements}

\bibliography{draft.bib}

\end{document}